\documentclass[reprint,onecolumn,a4paper,superscriptaddress,longbibliography,
amsmath,amssymb]{revtex4-2}

\usepackage{lineno}
\usepackage{mathtools}
\usepackage{amsmath}
\usepackage{graphicx}
\usepackage{booktabs,multirow}
\usepackage{braket}
\usepackage{textcomp}
\usepackage{yfonts}
\usepackage{comment}
\usepackage{xcolor}

\usepackage[normalem]{ulem}

\usepackage{geometry}
\usepackage[%
  bookmarks=true,
  colorlinks,
  linkcolor=blue,
  urlcolor=blue,
  citecolor=blue,
  plainpages=false,
  pdfpagelabels,
  final,
  breaklinks=true
]{hyperref}

\begin{document}

\newcommand{\new}[1]{\textcolor{red}{#1}}
\newcommand{\blue}[1]{\textcolor{blue}{#1}}
\newcommand{\so}[1]{\textcolor{green}{\sout{#1}}}

\newcommand{\pd}[2]{\frac{\partial #1}{\partial #2}} 
\newcommand{\td}[2]{\frac{d #1}{d #2}} 

\newcommand{\bs}{\boldsymbol}
\newcommand{\bt}{\textbf}
\newcommand{\sech}{\text{sech}} \newcommand{\erfc}{\text{erfc}}
\newcommand{\bse}{\begin{subequations}} \newcommand{\ese}{\end{subequations}}
\newcommand{\im}{\text{i}} \newcommand{\ud}[0]{\mathrm{d}}
\newcommand{\norm}[1]{\left\lVert#1\right\rVert}
\newcommand{\ve}{\varepsilon}
\newcommand{\op}{\widehat}

\graphicspath{{../figures/},{figures/}}
\allowdisplaybreaks

\title{Ray and caustic structure of Ince-Gauss beams
}

\author{R. Guti\'errez-Cuevas}
\email{rodrigo.gutierrez-cuevas@espci.fr}
\affiliation{Institut Langevin, ESPCI Paris, Université PSL, CNRS, 
75005 Paris, France}
\affiliation{Aix Marseille Univ, CNRS, Centrale Marseille, Institut Fresnel,
 UMR 7249, 13397 Marseille Cedex 20, France}
\author{M.~R.~Dennis}
\email{m.r.dennis@bham.ac.uk}
\affiliation{School of Physics and Astronomy, University of Birmingham, 
Birmingham B15 2TT, UK}
\author{M.~A.~Alonso}
\email{miguel.alonso@fresnel.fr}
\affiliation{Aix Marseille Univ, CNRS, Centrale Marseille, Institut Fresnel, 
UMR 7249, 13397 Marseille Cedex 20, France}
\affiliation{The Institute of Optics, University of Rochester, Rochester, NY 14627, USA}
\affiliation{Laboratory for Laser Energetics, University of 
Rochester, Rochester, NY 14627, USA}




\date{\today}

\begin{abstract}
The Ince-Gauss beams, separable in elliptic coordinates, are studied through a ray-optical approach. 
Their 
ray structure can be represented over
a ray-Poincar\'e sphere by generalized Viviani curves (intersections of a cylinder and a sphere). This representation shows two topologically different regimes, in which the curve is composed of one or two loops. The overall beam shape is described by the ray caustics that delimit the beams' bright regions. These caustics are inferred from the generalized Viviani curve through a geometric procedure that reveals connections with other physical systems and geometrical constructions.  Depending on the regime, the caustics are composed either of two confocal ellipses or of segments of an ellipse and a hyperbola that are confocal. The weighting of the rays is shown to follow the
two-mode meanfield Gross-Pitaevskii equations, which can be mapped to the equation of a simple pendulum. Finally, it is shown that the wave field can be accurately estimated from the ray description.
\end{abstract}

\maketitle

\section{Introduction}

The paraxial wave equation accepts many types of solutions,  some of which are separable in a given coordinate system
\cite{boyer1975liea,siegman1986lasers,siviloglou2007observation,
bandres2008accelerating,levy2016light,dennis2013propagation,
gutierrez-cuevas2017polynomials,gutierrez-cuevas2017complete}. A particular case is that of
beams that take the form of a Gaussian times a polynomial involving the corresponding separation variables  \cite{boyer1975liea,boyer1975lie,siegman1986lasers}. The Hermite-Gauss (HG)
and Laguerre-Gauss (LG) beams, for example, are separable in Cartesian and polar coordinates, respectively, and are used in a range of applications including quantum information,
telecommunications and imaging
\cite{andrews2008structured,berkhout2010efficient,zhou2017sorting,gu2018gouy,
tsang2016quantum,yao2011orbital}. Given the simplicity of the closed-form expressions for these beams in the wave regime, only a handful of studies consider them in terms of a ray-optical picture \cite{berry2008exact,alonso2017ray,dennis2019gaussian}.

The Ince-Gauss (IG) beams
\cite{arscott1964periodic,boyer1975lie,bandres2004incegaussiana,
bandres2004incegaussian,yaoli2020classically,sakpal2018stability,
gutierrezcuevas2023exactly}, separable in elliptic coordinates, have received significant attention in the last couple of decades, finding also many applications 
\cite{dennis2006rows,woerdemann2011optical,krenn2013entangled,plick2013quantum}.
Elliptical coordinates 
require specifying 
the separation $2f$ between its two foci. 
Let the radial $\xi \in[0,\infty)$ and angular $\nu \in[0,2\pi)$ elliptic variables be defined as $\bt r=(x,y)=(f\cosh \xi \cos \nu,f\sinh \xi \sin \nu)$. Within a region of some length scale $w_0$ around the origin, these elliptic coordinates  tend to the Cartesian coordinates as $f\to\infty$, while for $f\to0$ the contours of this system reduce to radial lines and circles as in polar coordinates. 
IG beams are then a broader family of beams that include both HG and LG beams as its two limiting
cases. The specific shape of an IG beam depends on the dimensionless parameter $\ve=2f^2/w_0^2$, where $w_0$ is the fundemantal width of the Gaussian. For any fixed $\ve$, IG beams form a complete orthonormal set in terms of their total order $N$, parity
p ($=$ e for even modes and o for odd modes), and index $\mu$, which runs from
0 (1) to $N$ in steps of two for $\rm p=\text{e}$ (o). 
At the waist plane $z=0$ these beams are defined (to within an unimportant normalization factor) in terms of the Ince polynomials $K_{N,\mu}^{(\text p)}$ as
\begin{align}
    \text{IG}_{N,\mu}^{(\text p)}(\bt r ;\ve)\propto 
    K_{N,\mu}^{(\text p)}(\im \xi , \ve)K_{N,\mu}^{(\text p)}(\nu , \ve) e^{-r^2/{w_0}^2}.
\end{align} 

\begin{figure}
    \centering
    \includegraphics[width=.95\linewidth]{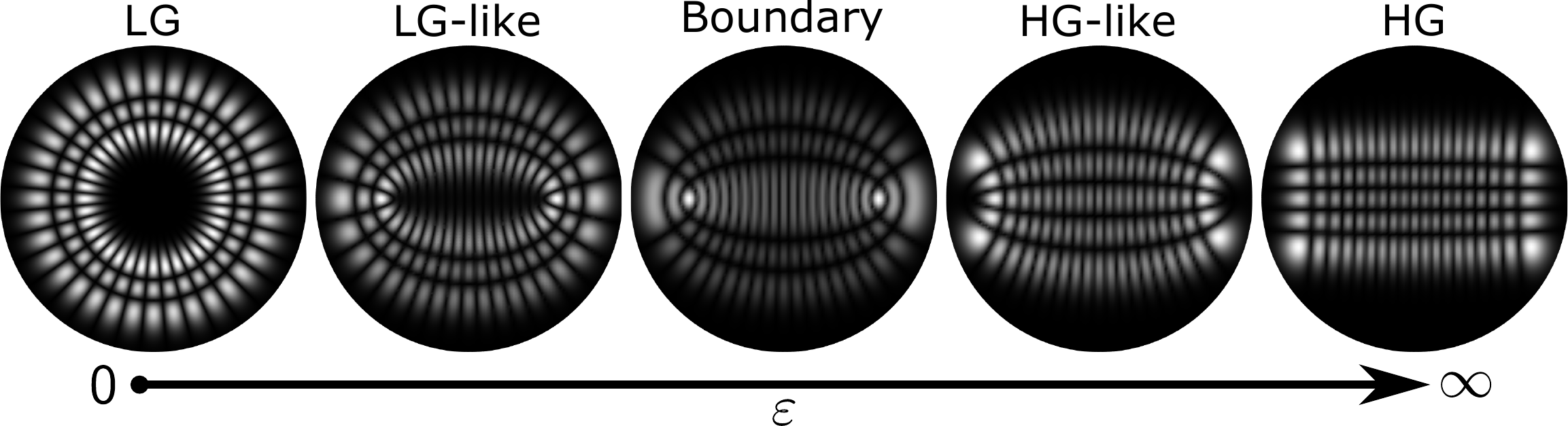}
    \caption{\label{fig:IG} 
    Intensity
    distribution for the even IG beam with total order $N=22$ and $\mu =18$ as
    $\ve$ varies from $0$ to $\infty$. At the extremal
    values $\ve=0$ and $\infty$, the IG beams reduce to real LG and
    HG beams, respectively. At intermediate values they take the
    form of either deformed LG-like or HG-like beams when $\ve$ is below or
    above a given boundary. }
\end{figure}

The variation
of an IG beam profile with $\ve$ (for fixed values of the beam's indices) is shown in Fig.~\ref{fig:IG}. Note that for $\ve\to0$ the beam reduces to a \emph{real} LG beam with $(N-\mu)/2$  radial nodes and $2\mu$ azimuthal nodes (that is, $\mu$ azimuthal field oscillations). At the other limit, for $\ve\to\infty$ the beam reduces to a HG beam with $\mu$ and $N-\mu$ nodes in each of the Cartesian directions. As $\ve$ varies between these limits the profile undergoes a transition in which 
two regimes can be recognized: the \emph{LG-like regime} where the beam resembles a deformed LG
beam since it is confined between two concentric loops, and the \emph{HG-like regime} where the beam resembles a deformed HG beam since it is confined within a quadrangle with curved sides. The distinction between these two regimes becomes more marked for modes with higher $N$. Note that, in all cases, the beams carry no net intrinsic orbital angular momentum (OAM), and hence the limit $\ve\to0$ corresponds to real LG beams and not ones presenting a phase vortex. 

The separable Gaussian solutions discussed so far are part of a larger set of paraxial beams known as structured Gaussian (SG) beams \cite{dennis2017swings,alonso2017ray,malhotra2018measuring,dennis2019gaussian,
gutierrez-cuevas2019generalized,
gutierrez-cuevas2020modal,gutierrezcuevas2020platonic}. SG beams preserve their intensity profile under propagation (up to a hyperbolic scaling), and are the modes of 
stable resonant cavities with spherical mirrors
\cite{siegman1986lasers,bandres2004incegaussian,schwarz2004observation}.
These cavities happen to be
mathematically analogous to a two-dimensional rotationally symmetric quantum harmonic oscillator, and thus the functional form of SG beams at their waist plane also describes the eigenmodes of this oscillator. The quantization of the cavity modes is analogous to the energy quantization of the oscillator modes, and given the systems' symmetry they both depend on only one index: the total order $N$. All modes with equal $N$ are then degenerate, explaining why a range of mode shapes (including HG, LG, IG for any $\ve$ and many others, all with any orientation) are possible.  The degenerate subspace of modes with given $N$ can be mapped onto
a spin system with SU(2) symmetry by using
Schwinger's coupled oscillator model \cite{sakurai2010modern}. This connection has allowed the use of various spherical representations to describe SG beams and their changes under specific modal transformations, as well as the phases they can accumulate
\cite{padgett1999poincare,calvo2005wigner,alonso2017ray,gutierrez-cuevas2020modal}. 

The degeneracy of the mirror cavity can be broken by inducing perturbations, such as small amounts of well-chosen aberrations. It was shown recently that IG modes with a given $\ve$ can be selected by adding small amounts of astigmatism and spherical aberration \cite{gutierrezcuevas2023exactly}, their ratio being proportional to $\ve$. In particular, a small amount of only spherical aberration selects LG modes, while a small amount of astigmatism alone leads to HG modes aligned with the symmetry axes of the aberration. Furthermore, this aberrated cavity system was shown to be
mathematically analogous to a Bose-Hubbard dimer, a widely used model in
condensed matter physics \cite{gati2007bosonic,gutierrezcuevas2023exactly}.

In this work, we take a fresh look at IG beams from a ray-optical perspective. Rays, and their envelopes known as caustics, have been used to explain intuitively the behavior of wave fields
\cite{berry1980iv,nye1999natural,kravtsov1999caustics,alonso2002stable,alonso2009rays,alonso2017ray,
malhotra2018measuring,gutierrezcuevas2020platonic}. 
When applied to SG beams, the ray model 
can explain both their ``self-healing'' behavior, as well as the Gouy phase accumulated around the focal region and the geometric phase accumulated during a sequence of mode transformations  \cite{alonso2017ray,malhotra2018measuring}. Rays can also be used to design new SG beams \cite{alonso2017ray,shen2020structured}, and to describe even SG beams that can be deemed as ``the least ray-like'' 
\cite{gutierrezcuevas2020platonic}. 
Here we show that a ray-based description provides important insights into the shape of IG beams, revealing a surprising amount of geometry that connects them to several other physical phenomena. In particular, this ray treatment is naturally compatible with the cavity aberrations mentioned earlier, and hence clarifies the two regimes (HG-like and LG-like) from the shape of the resulting caustics, as well as the boundary between these regimes, which can be understood as a ray-optical topological transition. We also show that the ray description is essentially complete, since it is sufficient to calculate accurate wave field estimates.

\section{Ray formalism for SG beams}

\begin{figure}
    \centering
    \includegraphics[width=.85\linewidth]{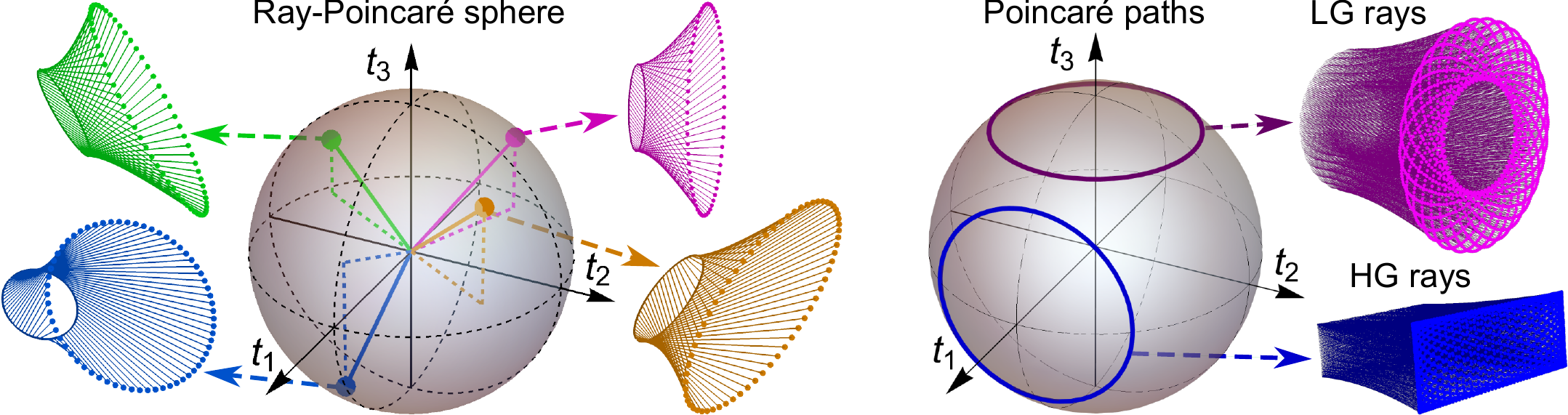}
    \caption{\label{fig:erfs} 
    (left) Ray-Poincar\'e sphere representation for elliptic
    families of rays (EFRs). Several EFRs are shown corresponding to different 
    points over the ray-Poincar\'e sphere.  
    (right) Examples of Poincar\'e paths for \emph{vortex} LG and HG beams.
    }
\end{figure}

In the paraxial ray formalism, at a given propagation distance $z$, say $z=0$, each ray can be identified by its transverse coordinates $\bt Q=(Q_x,Q_y)$ and its transverse
direction cosine vector $\bt P=(P_x,P_y)$. These quantities can be used to constitute a
four-dimensional phase space $(Q_x,Q_y,P_x,P_y)$, where each point represents a possible ray. An
optical beam is represented by a two-parameter family of rays that defines a
surface (referred to as the Lagrange manifold) within this phase space \cite{kravtsov1999caustics,alonso2009rays}. Typically, for well-localized beams, this manifold is topologically a torus, so that the two parameters, say $\tau$ and $\eta$, are periodic.  For SG beams, the requirement of self similarity means that the cross-section of the ray family must maintain its overall shape up to a global scaling, and as shown elsewhere \cite{alonso2017ray,dennis2019gaussian} this imposes a specific
dependence for $\bt Q$ and $\bt P$ on one of the two parameters, say $\tau$, so that at 
$z=0$ they satisfy
\begin{align}\label{eq:QandP}
    \frac{1}{w_0} \bt Q(\eta ,\tau) +\im \frac{kw_0}{2}  \bt P(\eta ,\tau)= 
        & \sqrt{N+1} \, \bt v(\eta)e^{-\im \tau},
\end{align}
where $\bt v$ is a two-dimensional complex vector with unit norm.
The normalization of $\bt v$
implies that 
\begin{align}
    \label{eq:qpcons}
    \frac{1}{w_0^2}\norm{\bt Q}^2 + \frac{k^2 w_0^2}{4} \norm{\bt P}^2 = N+1 .
\end{align}
This expression is equivalent to the equation for the conservation of energy of a
two-dimensional harmonic oscillator, and it restricts the region of the 4D phase space
occupied by the Lagrange manifold. Note that, for fixed $\eta$, both $\bt Q$ and $\bt P$ describe ellipses as
$\tau$ varies, just like the time evolution for the position
and momentum of a harmonic oscillator. Upon propagation, each of these elliptic families of
rays (EFR) describes a ruled hyperboloid with an elliptic cross-section that
satisfies the property of self similarity, as shown in Fig.~\ref{fig:erfs}: the elliptical cross-sections get larger upon propagation, but they retain their ellipticity and orientation.

An EFR is fully determined by the normalized complex two-vector $\bt v$. Since the vector is normalized and its global phase is irrelevant, it can be parametrized
as 
\begin{align}
    \label{eq:jonesv}
    \bt v(\eta) = 
    \cos \left[ \theta(\eta) / 2 \right] e^{-\im \frac{ \phi(\eta)}{2}} \bs \epsilon_+ 
    + \sin \left[ \theta(\eta) /2\right] e^{\im \frac{ \phi(\eta)}{2}} \bs \epsilon_- ,
    \qquad \bs \epsilon_\pm =(\hat{\bt x} \pm \im \hat{\bt y})/2^{1/2},
\end{align}
with $\theta$ and $\phi$ being the colatitude and longitude angles of a sphere,
respectively. This vector is then mathematically equivalent to the Jones vectors used to represent
polarization states as points on the surface of a Poincar\'e sphere. It is then possible to represent any EFR as
a point on the surface of ray-Poincar\'e sphere as shown in Fig.~\ref{fig:erfs}.
Like for polarization, the angle $\theta$ determines the ellipticity and handedness of the EFR, while $\phi$ determines its orientation. These angles are
functions of the second parameter, $\eta$, so that an SG beam is described by a
path over the surface of the sphere, referred to here as the Poincar\'e path
(PP). Because this second parameter is typically periodic, the PP is typically a closed path.

The analogy between the ray structure for SG beams and polarization can be
extended by introducing analogs of the Stokes parameters, which
define a Cartesian coordinate system for the ray-Poincar\'e sphere. These are the Fradkin-Stokes parameters, defined as
\cite{dennis2017swings,gutierrezcuevas2023exactly}
\bse
\label{eq:FS}
\begin{align}
    T_1 = & \frac{1}{2w^2_0}(Q_x^2-Q_y^2) + \frac{k^2w_0^2}{8}(P_x^2 - P_y^2),\\
    T_2 = & \frac{1}{w_0^2}Q_xQ_y + \frac{k^2w^2}{4} P_xP_y ,\\
    T_3 = & \frac{k}{2} (Q_xP_y - Q_yP_x).
\end{align}
\ese
For each ray, these parameters are invariant under propagation in free space, with $T_3$ being
proportional to the skew invariant, which is a measure of the OAM of each ray.
Further, 
it can be shown that these parameters are constant 
also over all rays
within a given EFR as defined by the expression in Eq.~(\ref{eq:QandP}) for fixed
$\eta$ and varying $\tau$. In fact, an alternative definition for the EFR is as the
set of all rays with given values for all three $T_n$. It can be shown that 
Eq.~(\ref{eq:qpcons}) can be written as 
\begin{align}
    T_1^2+T_2^2+T_3^2 = \left( \frac{N+1}{2} \right)^2,
\end{align}
so these parameters are indeed constrained to a spherical surface in the space $(T_1,T_2,T_3)$. Because the
Poincar\'e sphere is often defined as a unit sphere, it is convenient to use the
normalized quantities $t_j = 2T_j/(N+1)$, so that 
 \begin{align} \label{eq:sphere}
     t_1^2 + t_2^2 +t_3^2 =1. 
 \end{align}

The ray families of the most common SG beams are simply related to the
parameters $T_j$ (and hence $t_j$). For example, LG beams have definite OAM, and hence their ray-optical representation is 
composed of only rays with given $t_3$. Their PP then
corresponds to the intersection of the sphere with a plane of constant $t_3$,
that is, to a horizontal circle over the sphere whose height determines the OAM as shown in Fig.~\ref{fig:erfs}.
A similar relation holds for HG beams oriented along the $x$ and $y$ axes, which
are composed instead of rays with given $t_1$, and whose PP are circles normal
to the $t_1$ axis (see Fig.~\ref{fig:erfs}). Ray families for which a linear combination of $t_1$ and
$t_2$ is constant correspond to HG beams with different orientations. More
generally, ray structures for which a linear combination of all three parameters
is constant, that is, where the PP is a circle with any orientation, correspond
to the so-called generalized Hermite-Laguerre-Gauss (HLG) beams
\cite{abramochkin2010generalized,dennis2017swings,dennis2019gaussian,
gutierrez-cuevas2019generalized}, which include the HG and LG beams as special
cases. 
Note that, except for these two special cases, HLG beams are not
separable in any coordinate system.

\section{Ray structure for the IG beams} \label{sec:IGrays}

\subsection{Poincar\'e path as a generalized Vivani curve}

\begin{figure}
    \centering
    \includegraphics[width=.95\linewidth]{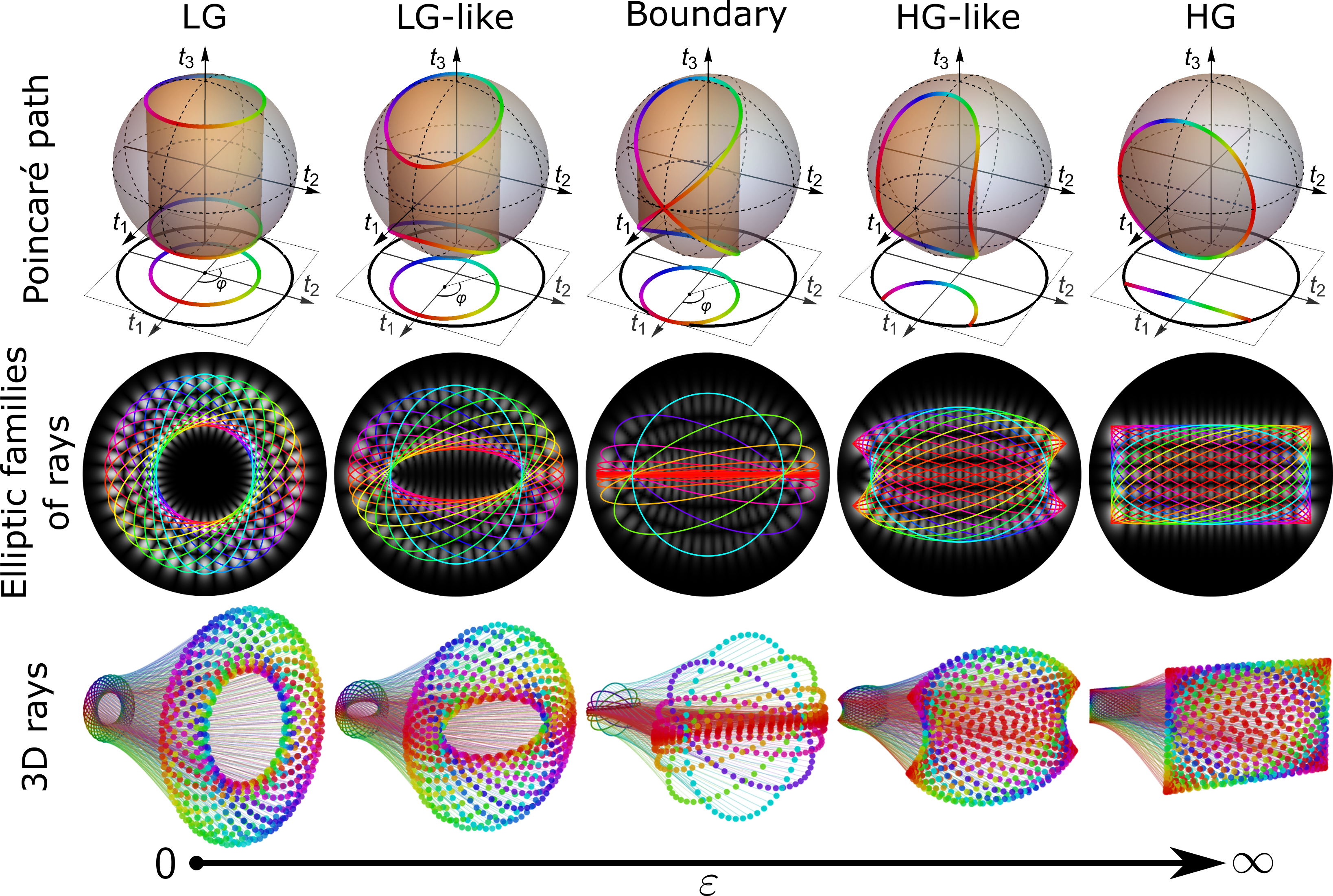}
    \caption{\label{fig:raysIG} Ray structure of an IG beam ($N=22$, even and
    $\mu=18$) as it transitions between the LG ($c=0$) and HG ($c \rightarrow
    \infty$) limits. (top) The intersection of the RPS with the surfaces
    defining the Poincar\'e path (PP) as their intersection along with the
    projection of the PP on the EPD where the angle $\varphi$ is defined.
    (middle) Elliptic families of rays color coded as the PP on the first row
    overlaid on top of the amplitude distribution for the corresponding beam. 
    (bottom) Three-dimensional propagation of the families of rays.}
\end{figure}

IG laser modes are known to result from aberrations or misalignment in a cavity
\cite{schwarz2004observation,gutierrezcuevas2023exactly}. In fact, before
IG beams were proposed, modes in aberrated cavities resembling IG modes were
observed experimentally and studied theoretically in terms of ellipsoidal
coordinates (beyond the paraxial regime) \cite{chao1974high}. We showed recently
\cite{gutierrezcuevas2023exactly} that $T_1$ can be associated with
the effect of a small amount of astigmatism in the cavity mirrors, whose
effect is precisely to break the symmetry in the $x$ and $y$ directions, while a
small amount of spherical aberration (which is nonlinear in ${\bf Q}$ and ${\bf
P}$) preserves rotational symmetry and can be associated instead with $T_3^2$,
where the square is consistent with the fact that the aberration does not have
an intrinsic skewness. 

As shown in Ref.~\cite{gutierrezcuevas2023exactly}, IG modes result when both aberrations are present. These modes are therefore associated with ray families for which a linear combination of $T_3^2$ and $T_1$ equals some constant, where $\ve$ determines the ratio between these aberrations:
\begin{align} \label{eq:UNNOR}
    T_3^2 +\frac{\ve}{2} T_1  = 
        \frac{a}{4}.
\end{align}
That is, unlike HLG beams, these
modes involve quadratic combinations of the Stokes-Fradkin parameters. This relation can be written in terms of the normalized parameters as
\begin{align} \label{eq:rayeigI}
t_3^2 + \frac{\ve}{N+1}t_1=\frac{a}{(N+1)^2}.
\end{align}
This equation defines a parabolic cylinder parallel to the $t_2$ axis, whose intersection with the unit sphere corresponds to the PP.
However, the resulting curve is easier to understand by using Eq.~(\ref{eq:sphere}) to
eliminate $t_3$ from Eq.~\eqref{eq:rayeigI}, giving
\begin{align} \label{eq:cylin}
\left(t_1 -c \right)^2 +t_2^2 = R^2,
\end{align}
which is the equation of a circular cylinder of radius $R$, with an axis parallel to the $t_3$ axis and that crosses the $t_1$ axis at
$c$, with
\begin{align}
c=\frac{\ve}{2(N+1)},\quad \text{and} \quad R=\sqrt{ 1  +\left[\frac{\ve}{2(N+1)} \right]^2 - \frac{a}{(N+1)^2} }.
\end{align}
The PP for IG modes then corresponds to the intersection of the ray-Poincar\'e sphere with this cylinder, as shown in Fig.~\ref{fig:raysIG}. Note that the condition $-1<c-R<1$ must be satisfied for an intersection between the cylinder and the sphere to
exist. 
For the specific case of $c=R=1/2$, this intersection corresponds to the Viviani curve (or Viviani window), proposed by seventeenth century mathematician Vincenzo Viviani \cite{abbena2017modern}. 
More generally, intersections of a cylinder with a sphere are
referred to as generalized Viviani curves, or also as Euclidic spherical ellipses since the sum of the distances
to two focal points on the surface of the sphere equals a constant \cite{graefe2014bosehubbard}. Generalized Viviani curves appear naturally in the description of a range of physical phenomena, including the Bose-Hubbard dimer \cite{graefe2014bosehubbard}, which as discussed earlier presents mathematical analogies with the aberrated cavity \cite{gutierrezcuevas2023exactly}. In optics, generalized Viviani curves describe polarization evolution in nonlinear birefringent fibers (also analogous to the Bose-Hubbard dimer) \cite{morales2017polarization}, and the subset of the generalized Viviani curves that are eight-shaped (with $c+R=1$) describe the polarization evolution for linearly polarized light traversing a rotating waveplate, or conversely the polarization states explored by a rotating-waveplate polarimeter \cite{azzam2000poincare}.   

We can now see that the regimes for the shape of IG
beams in Fig.~\ref{fig:raysIG} correspond to the different types of generalized Viviani curves for the PP: 
\begin{enumerate}
\item For $c+R<1$ (that is, for $\ve<a/(N+1)$) the PP consists of two disjointed loops, and this defines the LG-like
regime, since the PP resembles a deformed version of that for a real LG beam (two horizontal circles that are mirror images of each other with respect to the $t_3=0$ plane).
\item For $c+R=1$ (that is, for $\ve=a/(N+1)$) the PP is an eight-shaped curve, which defines the boundary between the two regimes. 
\item For $c+R>1$ (that is, for $\ve>a/(N+1)$) the PP consists of a single loop, and this defines the HG-like regime
since the PP resembles a deformed version of that for HG beams (a single vertical circle).
\end{enumerate}
That is, in this ray-optical description, the transition between the LG-like
and the HG-like regimes for IG beams corresponds to a topological transition for the ray family: in the LG-regime, the ray family is actually composed of two separate ray families (two phase-space tori) that are mirror images of each other. As $\ve$ increases these two families deform until they share an EFR for $\ve=a/(N+1)$, and they then merge into a single continuous ray family (one phase-space torus) in the HG-like regime. Note that while the boundary is sharp in the ray regime, it becomes blurry when we take into account the wave nature of light. That is, for values of $\ve$ near the transition there can be appreciable tunneling effects between the two separate parts of the PP that approach each other. 
The shape correspondence between the ray and wave pictures is shown in Fig.~\ref{fig:raysIG}.

\subsection{Caustic structure in terms of Keplerian and harmonic oscillator trajectories}

\begin{figure}
    \centering
    \includegraphics[width=.75\linewidth]{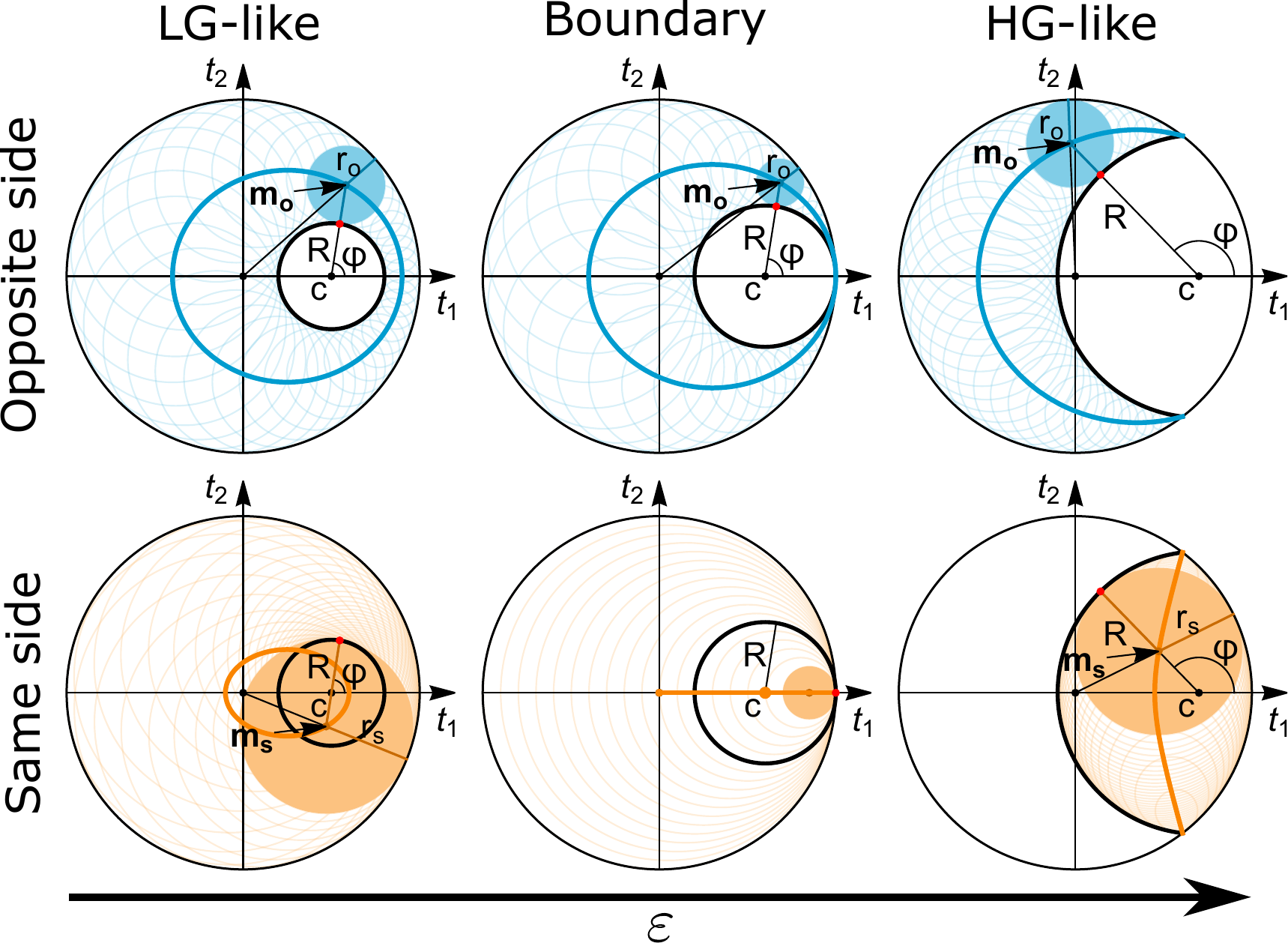}
    \caption{\label{fig:medial} 
    Construction of the medial axes from the PP projection for the two regimes and their boundary. (top) Medial axis (thick blue curve) traced for varying $\varphi$ by the point $\bt m_{\rm o}$ which is the center of the bitangent circle (blue) of radius $r_{\rm o}$ that touches both the edge of the unit disk and the PP projection (thick black circle with radius $R$) at the point (red dot) corresponding to a given angle $\varphi$. Note that the PP projection's center and $\bt m_{\rm o}$ are on opposite sides of the tangent point to the PP projection. (bottom) Same construction (with orange instead of blue) for the medial point $\bt m_{\rm s}$, which is on the same side of the tangent point of the PP projection as this curve's center.
     A more detailed view of this construction is shown in \textcolor{blue}{Media 1-4} for the different regimes.}
\end{figure}

\begin{figure}
    \centering
    \includegraphics[width=.95\linewidth]{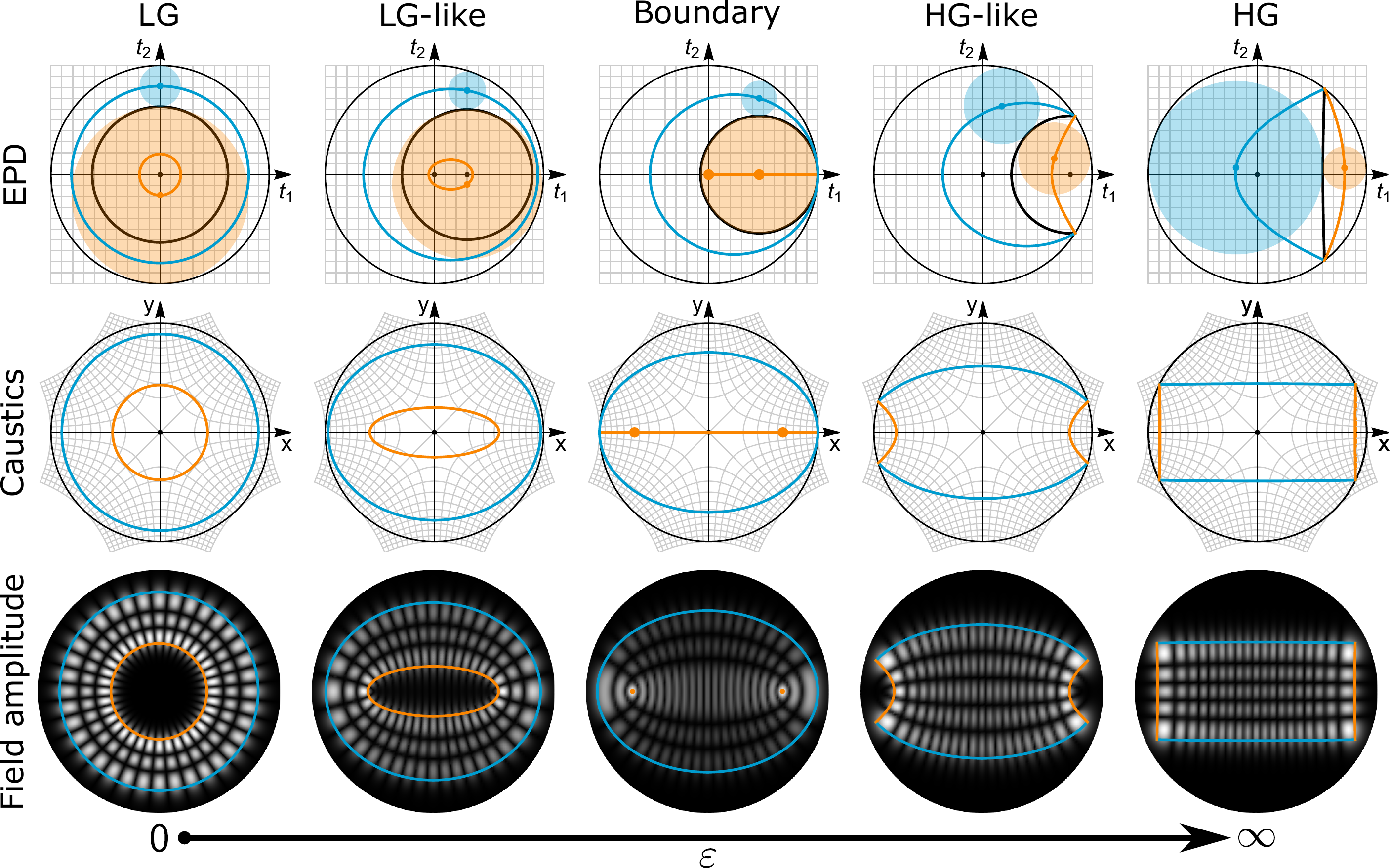}
    \caption{\label{fig:caustics} Construction of the caustics from the PP as
    they transition between the LG ($c=0$) and HG ($c \rightarrow \infty$)
    limits. (top) Equatorial Poincar\'e disk (EPD) with the projection of the PP
    (inner black curve) and the two medial axes (blue and orange curves) with
    examples of the bitangent circles used to construct them, (middle) the caustics
    obtained via the conformal mapping of the medial axes and (bottom) the
    caustics overlaid on top of amplitude distribution of the corresponding
    beam. }
\end{figure}

As we can see from the bottom row in Fig.~\ref{fig:raysIG}, all rays are restricted to a region of space confined by an outer surface, and also by an inner surface for the case of LG-like beams. These surfaces correspond to what is known as caustics, which are 
the envelopes of the rays. Due to the larger density of rays in their vicinity (on one side), caustics are associated with the brightest regions of an optical field \cite{berry1980iv,nye1999natural,kravtsov1999caustics}. 
Since the PP encodes the two-parameter ray family,
the caustics can be inferred from it. The procedure for doing so turns out to be of a geometrical nature \cite{alonso2017ray}, particularly for the case of IG beams considered here. 
The coordinates $(t_1,t_2)$ of the PP are sufficient for the determination of the caustics, since the only extra information provided by $t_3$ is the sign of the rays' skewness, which does not affect where they form an envelope. As shown in the top row of Fig.~\ref{fig:raysIG}, the ray-Poincar\'e sphere can then be projected onto a unit disk, referred to here as the equatorial Poincar\'e disk (EPD). Since for IG beams the PP is the intersection of a vertical circular cylinder with the sphere, the PP projects onto a circle (or a circular segment) of radius $R$ centered at $(t_1,t_2)=(c,0)$. This circle can be parametrized as
\begin{align} \label{eq:cylphi}
(t_1,t_2) = (c+R\cos \varphi, R\sin \varphi),
\end{align}
where $\varphi$ is the angle measured from the positive $t_1$ axis (see Figs.~\ref{fig:medial} and \ref{fig:caustics}). In the LG-like regime
$\varphi$ traces a complete cycle while in the HG-like regime it is constrained to
$[\varphi_0,2\pi -\varphi_0]$ with
\begin{align}\label{eq:maxvarphi}
\varphi_0 = \arccos h \qquad \text{with} \qquad h=\frac{1-R^2-c^2}{2Rc}.
\end{align}

We now describe the geometric procedure for finding the caustics from the PP projection. As discussed in Ref.~\cite{alonso2017ray}, the first step is to find what is known as the \emph{medial axes} or \emph{topological skeleton} \cite{blum1967transformation} between  
the PP projection in Eq.~\eqref{eq:cylphi} and the unit circle (the edge of the EPD). 
%
Medial axes are curves that are equidistant to two other curves, and are defined as the loci of centers of what we call here bitangent circles, which are the circles that are tangent to the two original curves. This is shown in Fig.~\ref{fig:medial} for the case of interest here, in which the two original curves are the PP projection and the unit circle. For each point of the PP projection (corresponding to a given value of $\varphi$), there are two bitangent circles, one of radius $r_{\rm o}$ and whose center $\bt m_{\rm o}$ is on the opposite side of the PP projection as its center, $(c,0)$, and one of radius $r_{\rm s}$ and whose center $\bt m_{\rm s}$ is on the same side of the curve as $(c,0)$. 
Because the PP projection is also a circle, the distance between its center $(c,0)$ and each of the two medial points is just the sum or difference of $R$ and the radius of the corresponding bitangent circle, 
namely  
\begin{align}
\label{eq:other}
    \norm{\bt m_{\rm o} - (c,0)}=r_{\rm o}+R,\qquad \norm{\bt m_{\rm s} - (c,0)}=|R-r_{\rm s}|.
\end{align}
Similarly, because both bitangent circles touch the unit circle, the distance between the origin and the medial points is the difference between unity (the radius of the unit circle) and the corresponding radius, that is, 
\begin{align}
\label{eq:same}
    \norm{\bt m_{\rm o}}=1-r_{\rm o},\qquad\norm{\bt m_{\rm s}}=1-r_{\rm s}.
\end{align}
The equations for the medial axes are obtained by eliminating the radii $r_{\rm o}$ and $r_{\rm s}$ from these relations, and we now consider each medial axis separately. 

For the medial axis traced by $\bt m_{\rm o}$ the elimination of $r_{\rm o}$ leads in all cases to the equation
\begin{align}
\norm {\bt m_{\rm o} } +\norm{\bt m_{\rm o} - (c,0)} =  1 + R,
\end{align}
which describes an ellipse with major axis $1+R$ and whose foci are the origin and $(c,0)$. This construction is shown in the top row of Fig.~\ref{fig:medial} and on the left side of \textcolor{blue}{Media 1-4}. Note that in the LG-like regime the medial axis traces the complete ellipse, and the geometry of this construction is related of what is known as a Steiner chain \cite{ogilvy1990excursions}, a remarkable geometric result by 19th-century Swiss mathematician Jakob Steiner. The ellipse is also traced completely in the boundary case, for which the construction resembles instead a Pappus chain \cite{ogilvy1990excursions}, described by Pappus of Alexandria in the third century. In the HG-like regime, on the other hand, only a fraction of the ellipse is traced by $\bt m_{\rm o}$, given the fact that the PP projection is an incomplete circle.

The equation for the second medial axis, traced by $\bt m_{\rm s}$, requires a different treatment for the two regimes, as well as for their boundary. In the LG-like regime we have that $r_{\rm s}>R$ and $R<1$, so that the right-hand side of the second expression in Eqs.~\eqref{eq:other} equals $r_{\rm s}-R$. The elimination of $r_{\rm s}$ from Eqs.~\eqref{eq:other} and \eqref{eq:same} then gives
\begin{align}
\norm {\bt m_{\rm s} } +\norm{\bt m_{\rm s} - (c,0)} = 1 -R,
\end{align}
which also describes a complete ellipse with foci at the origin and $(c,0)$ but with major axis $1-R$, as shown in the bottom-left panel of Fig.~\ref{fig:medial}) and the right panel of \textcolor{blue}{Media 1}. 
In the HG-like regime, on the other hand, because $R>r_{\rm s}$ the right-hand side of the second expression in Eqs.~\eqref{eq:other} equals $R-r_{\rm s}$, and the elimination of $r_{\rm s}$ gives instead
\begin{align}
\norm {\bt m_{\rm s}} -\norm{\bt m_{\rm s} - (c,0)} = 1 -R,
\end{align}
which describes a branch of a hyperbola with the same foci, as illustrated in the bottom-right panel of Fig.~\ref{fig:medial}) and the right panel of \textcolor{blue}{Media 3,4}. 
This hyperbolic branch presents different behavior for different values of $R$: it opens away from the origin for $R<1$ (as in the case shown in the bottom-right part of Fig.~\ref{fig:medial} and in \textcolor{blue}{Media 3}), it is a straight line for $R=1$, it opens towards the origin for $R>1$ (as shown in in \textcolor{blue}{Media 4}), and it becomes a parabola with focus at the origin in the limit $R\to\infty$ (corresponding to the HG case). In fact, in this latter limit, the ellipse traced by $\bt m_{\rm o}$ also becomes a parabola. 
At the boundary of the two regimes, for which $c+R=1$, the 
medial axis traced by $\bt m_{\rm s}$ 
undergoes a transition that makes it occupy the complete unit radial line corresponding to the positive $t_1$ axis, as shown in the bottom-center panel of Fig.~\ref{fig:medial}, because any circle centered on this line segment that is tangent to the unit disc at $t_1=1$ is also tangent to the PP projection and hence is a valid bitangent circle. Note, though, that two points along this straight medial axis are special, namely the origin and the point at $(c,0)$. When $\bt m_{\rm s}$ coincides with one of these points, $r_{\rm s}$ equals either unity or $R$, meaning that the bitangent circle is identical either to the edge of the EPD or the PP projection and hence is tangent to the corresponding curve not only at one point but all along its perimeter. This transition case is illustrated in the right-hand panel of \textcolor{blue}{Media 3}.

In summary, for both regimes the medial axes are conic sections with one focus at the origin of the EPD and the other at $(c,0)$. 
Therefore, geometrically, their shape coincides with that of Keplerian orbits for a potential centered at the origin. 
These orbits are closed (elliptic) and correspond to an attractive force for $\bt m_{\rm o}$ in all cases (although $\bt m_{\rm o}$ does not trace the complete orbit for $c+R>1$), and also for $\bt m_{\rm s}$ in the LG-like regime. For $\bt m_{\rm s}$ in the HG-like regime, the force is attractive if $R>1$ (although the orbit is not periodic), repulsive for $R<1$, and there is no force for $R=1$. In the limit $c\to\infty$ (and hence $R\to\infty$) corresponding to HG beams, both $\bt m_{\rm o}$ and $\bt m_{\rm s}$ trace parabolas corresponding to escape orbits, while in the opposite limit of $c\to0$ corresponding to LG beams, both $\bt m_{\rm o}$ and $\bt m_{\rm s}$ trace circular orbits. 

The medial axes can now be mapped directly onto the caustics
by using a square-root conformal map
\cite{alonso2017ray}. This is done
by associating a medial point $\bt m = (t_1,t_2)$ with the complex number
$
t_1 +\im t_2$ and then taking its square root:
\begin{align} \label{eq:sqrtmap}
Q^{(\rm c)}_x +\im Q^{(\rm c)}_y = \pm w_0 \sqrt{(N+1)
(t_1 +\im t_2)
}.
\end{align}
The resulting point $\bt Q^{(\rm c)} = (Q^{(\rm c)}_x , Q^{(\rm c)}_y)$ corresponds to a caustic point. Given the sign ambiguity of the square root, each medial axis point maps onto two caustic points, and therefore each of the two medial axes maps onto two segments of the caustic curves. 
Remarkably, the 
square-root conformal map is known to have the property of mapping Keplerian orbits onto the orbits of a particle in a
(repulsive or attractive) two-dimensional harmonic oscillator, which correspond to conic sections centered at the origin
\cite{mittag1992conformal}.  (Incidentally, the properties of this mapping when applied to conics make it also useful for studying the shape of light reflectors in non-imaging applications \cite{aleman-castaneda2019study}.) Since the medial axes of IG beams
correspond precisely to Keplerian orbits, their caustics take the shape of confocal conic sections with foci that coincide with those for the elliptical coordinate system, $(\pm f,0)$.
For beams in the LG-like regime, the caustics are two complete confocal ellipses, which in the limit of LG beams become two concentric circles. 
In the HG-like regime, on the other hand, the caustic structure is composed of two segments of an ellipse and two segments of a hyperbola, forming a deformed quadrangle, and in the limit of HG beams these segments become straight lines (since the square-root map transforms parabolas to straight lines). In the boundary case, the outer caustic is an ellipse and the inner one occupies the major axis of this ellipse, with two main points that coincide with the foci. In all cases, all rays are contained within these caustics.

\section{Semiclassical estimates}

\subsection{Wave reconstruction}
The wave behavior of an optical field can be reproduced faithfully from the ray description if an
appropriate semiclassical method is used. Since there is a closed-form expression for the IG beams, it is not necessary to rely on a semiclassical approximation to calculate them. However, it is still interesting to show that the ray picture is essentially sufficient to reconstruct these fields. Further, this asymptotic analysis helps illustrate some interesting aspects of the topological transition between the LG-like and the HG-like regimes, and helps elucidate a connection with yet another physical system. 

The semiclassical approach we use is a Gaussian summation method
\cite{zor1996globally,alonso2001using,alonso2002stable}, since it does not diverge at the caustics and was shown to reproduce the exact paraxial solutions for HG, LG
and HLG beams \cite{kay1999exact,kay2001exact,alonso2017ray,dennis2019gaussian}.
In this method, each ray is dressed  with a Gaussian contribution, 
and the estimate takes the form of an integral over $\tau$ and $\eta$. 
When applied to SG beams, the integral in $\tau$ over a period can be solved analytically, and 
the constant factors in Eq.~\eqref{eq:QandP} guarantee that the integrand is consistent at the endpoints of integration. The
semiclassical estimate then takes the form of an integral of contributions associated with each EFR that allows estimating a SG beam from its ray structure \cite{alonso2017ray,dennis2019gaussian}:
\begin{align}\label{eq:estU}
    U(\bt r) \approx \oint A(\eta) 
    \sqrt{\phi'(\eta)\sin \theta-\im \theta'(\eta)}\,
    U_N(\theta,\phi; \bt r)\, e^{\im (N+1) \Upsilon (\eta)}\,\ud \eta , 
\end{align}
where $A(\eta)$ is a weight factor for the contributions from each EFR,  primes denote derivatives with respect to $\eta$, $U_N$ is a mathematical analog of a spin coherent state
\cite{radcliffe1971some,arecchi1972atomic,perelomov1972coherent,
gutierrez-cuevas2020modal} defined as
\begin{align} \label{eq:coherent}
    U_N(\theta,\phi; \bt r)=
    \frac{(\bt v \cdot \bt v)^{N/2} }{w \sqrt{2^{N-1}\pi N! }}
    e^{-\frac{r^2}{w^2}}
    H_N\left( \frac{\sqrt 2 \bt v \cdot \bt r }{w\sqrt{\bt v \cdot \bt v} }\right),
\end{align}
with $H_N$ representing an $N$th order Hermite polynomial, and
\begin{align} \label{eq:upsi}
    \Upsilon(\eta) = \frac{1}{2} 
        \int_0^\eta \phi'(\bar \eta) \cos \theta(\bar \eta) \ud \bar \eta
\end{align}
gives a phase to each contribution that can be interpreted as a geometric phase, since it corresponds to half an area on the ray-Poincar\'e sphere. This integral can be performed analytically in terms of elliptic integral functions. (See appendix~\ref{app:upsi} for more information.)

\begin{figure}
\centering
\includegraphics[width=.95\linewidth]{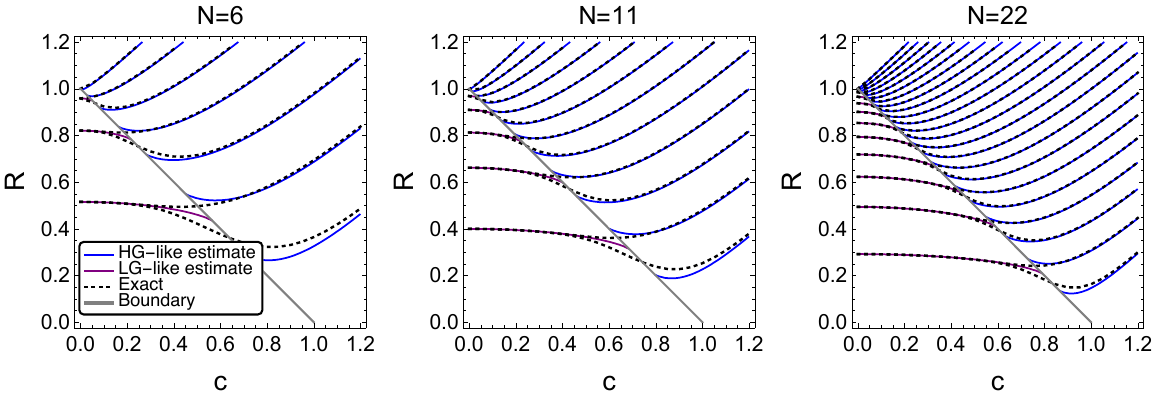}
\caption{\label{fig:eigen} Semiclassical and exact values for the radius $R$ for
the IG beams with $N=6$, $N=11$ and $N=22$ as a function of the center coordinate $c$. The different curves or pairs of curves correspond to different values of $\mu$.}
\end{figure}

While the periodicity of the integrand in $\tau$ is guaranteed by the factors in Eq.~\eqref{eq:QandP}, the periodicity of the integrand in $\eta$ imposes conditions on the phase contribution due to the geometric phase caused by the PP enclosing a solid angle $\Omega$ over the ray-Poincaré sphere. Because the square-root factor in Eq.~\eqref{eq:estU} accumulates a phase of $\pm\pi$ on closing the PP loop, the solid angle must then be an odd multiple of $2\pi/(N+1)$, 
\cite{alonso2017ray,dennis2019gaussian}
\begin{align} \label{eq:quant}
\Omega = \frac{2\pi}{N+1} (2s+1),
\end{align}
for integer $s$. 
This condition restricts the valid values for $R$ in Eq.~\eqref{eq:cylin}
needed to define the PP. However, recall that the number of closed loops composing the PP differs for the two regimes. For the LG-like regime there are two loops and each must satisfy the equation above, so the total subtended solid
angle must be an odd multiple of $4\pi/(N+1)$ (since the two loops enclose the same solid angle). On the other hand, in the HG-like regime the PP consists of a single loop so the solid angle 
must be an odd multiple of $2\pi/(N+1)$. This leads to a discontinuity at the ray-optical
boundary on the allowed values for $R$ for a given $c$; as shown in Fig.~\ref{fig:eigen} there are $N+1$ allowed values of $R$ for the HG-like regime, but only $\lfloor (N+1)/2\rfloor$ for the LG-like regime. 

\subsection{The Ince equation and its eigenvalues}
In the wave domain, IG beams are defined as eigenstates of what we call here the Ince equation, which corresponds to
the operator version of Eq.~\eqref{eq:UNNOR}:
\begin{align} \label{eq:eigenIG}
    \left( \op T_3^2 +\frac{\ve}{2}\op T_1 \right) \text{IG}_{N,\mu}^{(\text p)} = 
        \frac{a}{4}\text{IG}_{N,\mu}^{(\text p)},
\end{align}
where the operators $\op T_j$ are obtained from the expressions for $T_j$ in Eqs.~\eqref{eq:FS} by substituting 
$\bt P\to-\im k^{-1}(\partial_{x},\partial_{y})$. Note that $a$ corresponds now to an eigenvalue. The dotted lines in Fig.~\ref{fig:eigen} were calculated by using the values of $a$ found through the numerical solution of Eq.~\eqref{eq:eigenIG}. These curves are continuous at the boundary, and they merge by pairs as one transitions into the LG-like regime. This is because LG-like IG beams with equal indices but different parity are nearly degenerate, and semiclassically they are represented by the same pair of PP loops, the only difference being the relative phase between the two contributions. The discrepancy between the curves in Fig.~\ref{fig:eigen} for the semiclassical estimate and the result using the exact eigenvalue is due to the fact that the semiclassical approach neglects tunneling effects that take place near the boundary. A more accurate semiclassical
estimate would require considering asymptotic corrections to the amplitude term derived below, which would modify the phase consistency condition that leads to Eq.~\eqref{eq:quant}. This type of correction was used recently in the asymptotic estimation of eigenvalues for anharmonic potentials \cite{forbes2023asymptotic}. Nonetheless, away from the boundary between the two regimes, the curves produced by the semiclassical estimate are in good agreement with those calculated using the exact eigenvalues,  even for
moderate $N$.

\subsection{The semiclassical amplitude and a connection with the simple pendulum}

The remaining element for evaluating the semiclassical estimate in 
Eq.~\eqref{eq:estU} is to find the amplitude  $A$. The choice of this function, as long as its phase is constant and it does not vary abruptly, does not affect the fact that the resulting beam is self similar or that it has a shape similar to that of IG beams. However, the form for this function that makes the estimate approach closely the IG beams can be calculated from substituting Eq.~\eqref{eq:estU} into Eq.~\eqref{eq:eigenIG},
leading to an equation that can be separated into different orders of
$N$ (see appendix \ref{app:amp} for more detials). Unsurprisingly, the dominant term gives Eq.~(\ref{eq:UNNOR}), namely the PP. The
next term in the series leads to an equation for the amplitude $A(\eta)$, and hence to
the appropriate weighting for each EFR. The resulting equation for the weight takes a simple form when written in terms of the Cartesian parameters $t_j(\eta)$:
\begin{align}
    A^2 =  \frac{B}c \frac{t_3'}{t_2},
\end{align}
where $B$ is a constant of integration. 

\begin{figure}
\centering
\includegraphics[width=.75\linewidth]{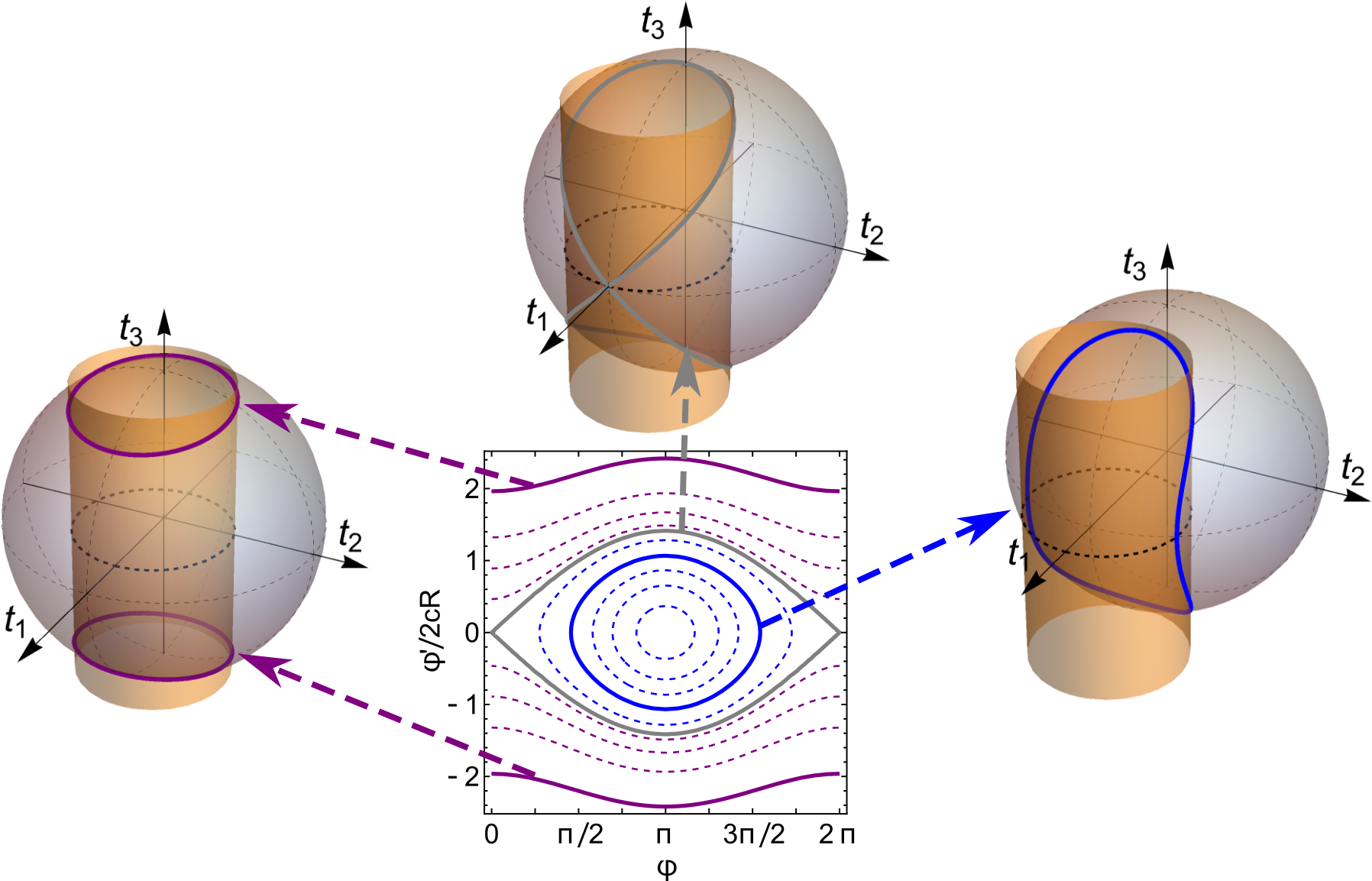}
\caption{\label{fig:phasepend} Mapping between the phase space curves of the pendulum and the PPs of IG beams on the ray-Poincar\'e sphere. 
}
\end{figure}

If we assume that the parametrization is
such that $A$ is constant, 
then $t_3'= A^2 t_2 / B$. For simplicity we set $A^2/B$ 
equal to unity. By combining this equation with the PP and the fact that the parameters $t_j$
lie on a unit sphere, we are led to the following system of first order
differential equations: 
\begin{align}
\label{eq:bloch}
    t_1' =  - t_2 t_3, \qquad
    t_2' =  t_1 t_3 -c t_3, \qquad
    t_3' =  c t_2. 
\end{align}
These equations are analogous to the two-mode Gross-Pitaevskii equations
describing the meanfield evolution of a Bose-Hubbard dimer
\cite{gati2007bosonic,graefe2014bosehubbard}. 
They lead to a
particularly simple form when written in terms of the angle $\varphi$ used to
parametrize the cylinder in Eq.~\eqref{eq:cylphi}:
\begin{align} \label{eq:pendulum}
\varphi''-  c R\sin \varphi= 0.
\end{align}
This is the equation for a simple pendulum where the parameter $\eta$
plays the role of time. This equation implies that the EFR contributions become more important the closest their point at the PP is to the equator (which correspond to points where the pendulum lingers). The two regimes for the IG beams then correspond to the two topologically
different regimes for the pendulum: full rotation corresponds to the LG-like regime and libration corresponds to the
HG-like regime. The analogous connection between the pendulum and the Bose-Hubbard dimer was pointed out by Graefe {\it et al.} \cite{graefe2014bosehubbard}. 
Further, the connection with the generalized Viviani curves becomes more intuitive if we use Eq.~\eqref{eq:cylphi} to write $t_3$ in terms of $\varphi$ as $t_3=\pm\sqrt{1-t_1^2-t_2^2}=\pm\sqrt{1-c^2-R^2-2cR\cos\varphi}$, and substitute this into the third relation in Eqs.~\eqref{eq:bloch}, which leads to the simple expression
\begin{align}
    \varphi'(\eta)=t_3.
\end{align}
That is, $t_3$ corresponds to the pendulum's velocity (or momentum, for a unit mass). One can then define a two-dimensional phase space $(\varphi,t_3)$ for the pendulum, in which its motion is described by a curve. However, given the possible periodicity of $\varphi$ one can roll up this phase space into a cylinder of radius $R$. The pendulum's phase space curve would then coincide with the PP for the IG beams, that is, with the intersection of the cylinder with a sphere as shown in Fig.~\ref{fig:phasepend}.

The fact that the rolling-up of the phase space of the pendulum falls on a sphere (or ellipsoid) is all the more surprising because the shape of the intersections of the cylinder with the sphere depends on two parameters, $c$ and $R$, while for a pendulum one considers typically that the orbits in phase space depend on a single parameter. Let us use as this parameter $h$ as defined in Eq.~\eqref{eq:maxvarphi}. The libration and rotation regimes for the pendulum then correspond to $h$ being smaller or larger than unity, respectively. It is easy to show that the equation for the PP, when unrolled onto the pendulum's $(\varphi,t_3)$ phase space, becomes $t_3^2=2cR(h-\cos\varphi)$, so that the shape of the curve depends only on $h$, and the factor of $2cR$ just provides a constant scaling for the pendulum's momentum that, amongst other things, guarantees that $|t_3|\le1$.

\subsection{Ray-based estimates}

\begin{figure}
\centering
\includegraphics[width=.95\linewidth]{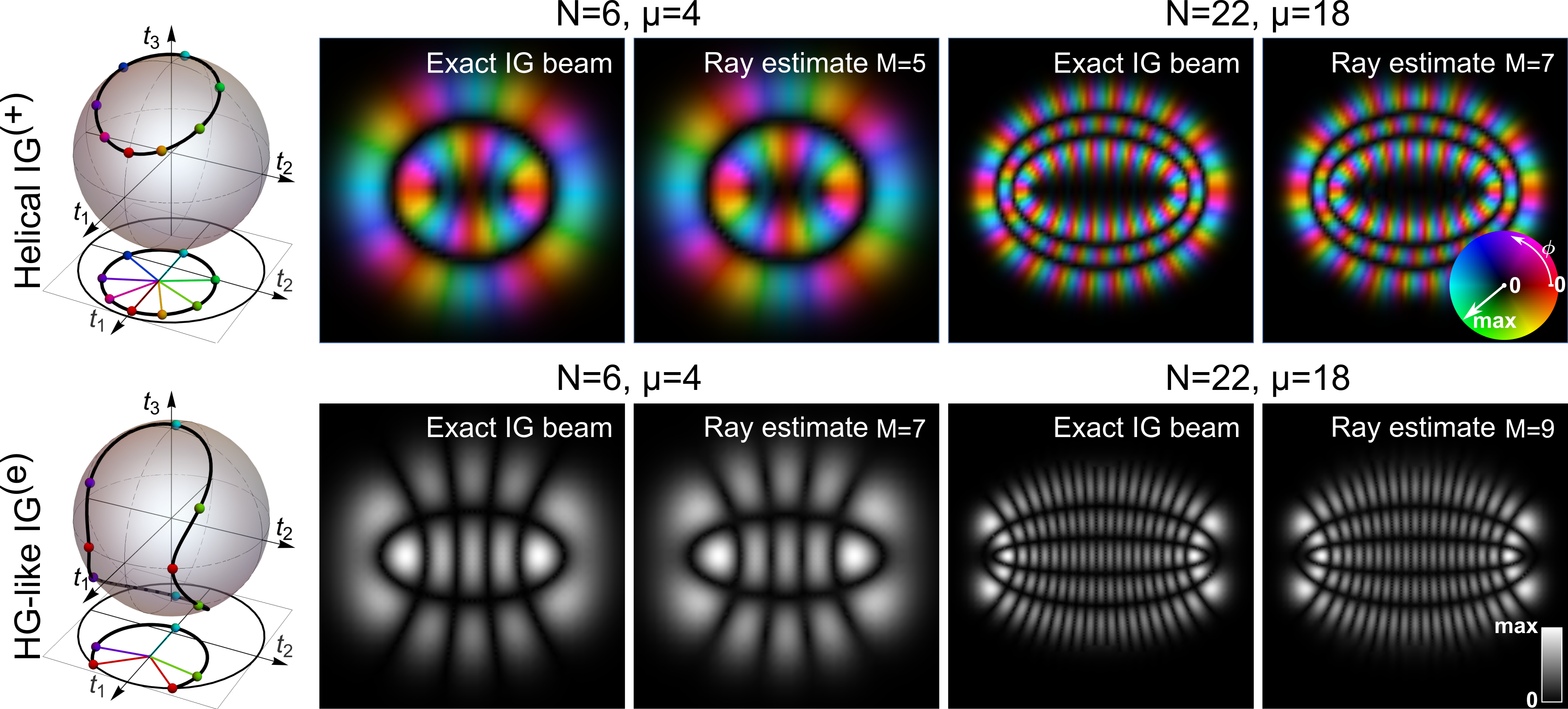}
\caption{\label{fig:rayest} Examples of sampling and comparison between 
the exact mode and its ray-based estimate for (first row) helical IG beams in
the LG-like regime with (second and third columns) $N=6$, $\mu=4$, $c=0.1$ and (fourth and fifth columns)
$N=22$, $\mu = 18$, $c=0.3$, and (second row) even IG beams in the HG-like regime with (second and third columns) $N=6$, $\mu=4$, $c=0.4$ and (fourth and fifth columns) $N=22$, $\mu = 18$, $c=0.7$. 
The number of terms $M$ was chosen as the smallest for which the estimate
is visually indistinguishable from the exact distribution. For the helical IG beams
the plots show the complex field with the phase encoded as hue, while for the HG-like
IG beams the plots show the amplitude distribution. The figures in the first column show an example of sampling for eight points at equal integrals in $\eta$. Note that their projection onto the $(t_1,t_2)$ plane is equivalent to snapshots at uniform time intervals of the evolution of a pendulum. 
}
\end{figure}

We conclude by testing the ray-based estimate for the IG
beams, to show that the ray families obtained recreate faithfully the wavefields. Since the integral in Eq.~\eqref{eq:estU} has no closed-form solution, we approximate it as a discrete sum of $M$ terms:
\begin{align} \label{eq:aproxestimate}
    U(\bt r) \approx \sum_{j=0}^{M-1}  
    \sqrt{\phi'(\eta_j)\sin \theta (\eta_j)-\im \theta'(\eta_j)}
    U_N(\theta (\eta_j),\phi (\eta_j); \bt r) e^{\im (N+1) \Upsilon (\eta_j)},
\end{align}
where we use the parametrization
defined by Eqs.~(\ref{eq:bloch}) and (\ref{eq:pendulum}) that makes $A=1$. The dependence of the
angles $\theta$ and $\phi$ on the parameter $\eta$ is determined by solving
these differential equations with appropriate initial conditions determined by
the PP. Setting $\eta_0 =0$ then $\eta_{j} = j\eta^{(\rm p)}/M $
where the pendulum's period is given by
\begin{align}
    \eta^{(\rm p)} =  
    \begin{cases}
    \int_0^{2\pi} \frac{\ud \eta}{\sqrt{1-c^2-R^2-2cR \cos \eta}}=4\frac{K(h)}{\sqrt{1-(R+c)^2}}, 
    & \text{if } c+R<1, \\
    \int_{\varphi_0}^{2\pi-\varphi_0} \frac{2\ud \eta}{\sqrt{1-c^2-R^2-2cR \cos \eta}}=8\frac{F(\varphi_0/2\,|\,h)-K(h)}{\sqrt{1-(R+c)^2}}, 
    & \text{if } c+R>1,
    \end{cases}
\end{align}
with $h=-4Rc/[1-(R+c)^2]$, and where $K$ and $F$ are, respectively, the complete and incomplete elliptic integrals of the first kind.

In the LG-like regime, 
each of the two loops provides a contribution, one being the complex conjugate of the other. 
However, the semiclassical theory does not constrain how these contributions are combined. Figure \ref{fig:rayest} shows the
field due to the northern PP loop, resulting in a  beam with the
same elliptical shape as the corresponding IG beam, but with nonzero net OAM. The southern loop would then have the opposite OAM. These are then estimates (when sufficiently far from the boundary) for what has been referred to as helical IG beams, which are
obtained by combining even and odd IG beams with the same $N$ and $\mu$ as
$\text{IG}^{(\pm)}_{N,\mu}= \text{IG}^{(\rm e)}_{N,\mu} \pm \im \text{IG}^{(\rm
o)}_{N,\mu}$ \cite{bandres2004incegaussian}.  
As shown in Fig.~\ref{fig:raysIG}, for both the upper and lower parts, the projections on the EPD of the sampling points used to compute the estimate provide us with snapshots of the evolution of a pendulum in full rotation. 
Real IG beams can be obtained
by adding both contributions either in phase or out of phase (that is, by taking the real or imaginary part of the contribution from one loop). In the HG-like regime, on the other hand, the PP is given by a single loop so that
Eq.~\eqref{eq:aproxestimate} directly provides a ray-based estimate for the
corresponding IG beam. Here, the projections on the EPD of the sampling points used to compute the estimate correspond to snapshots at equal temporal intervals of the libration evolution of the pendulum.
Note that it is not possible to obtain an estimate at the boundary between the two regimes because the quantization condition becomes
ill-defined.

%
\begin{figure}
\centering
\includegraphics[width=.95\linewidth]{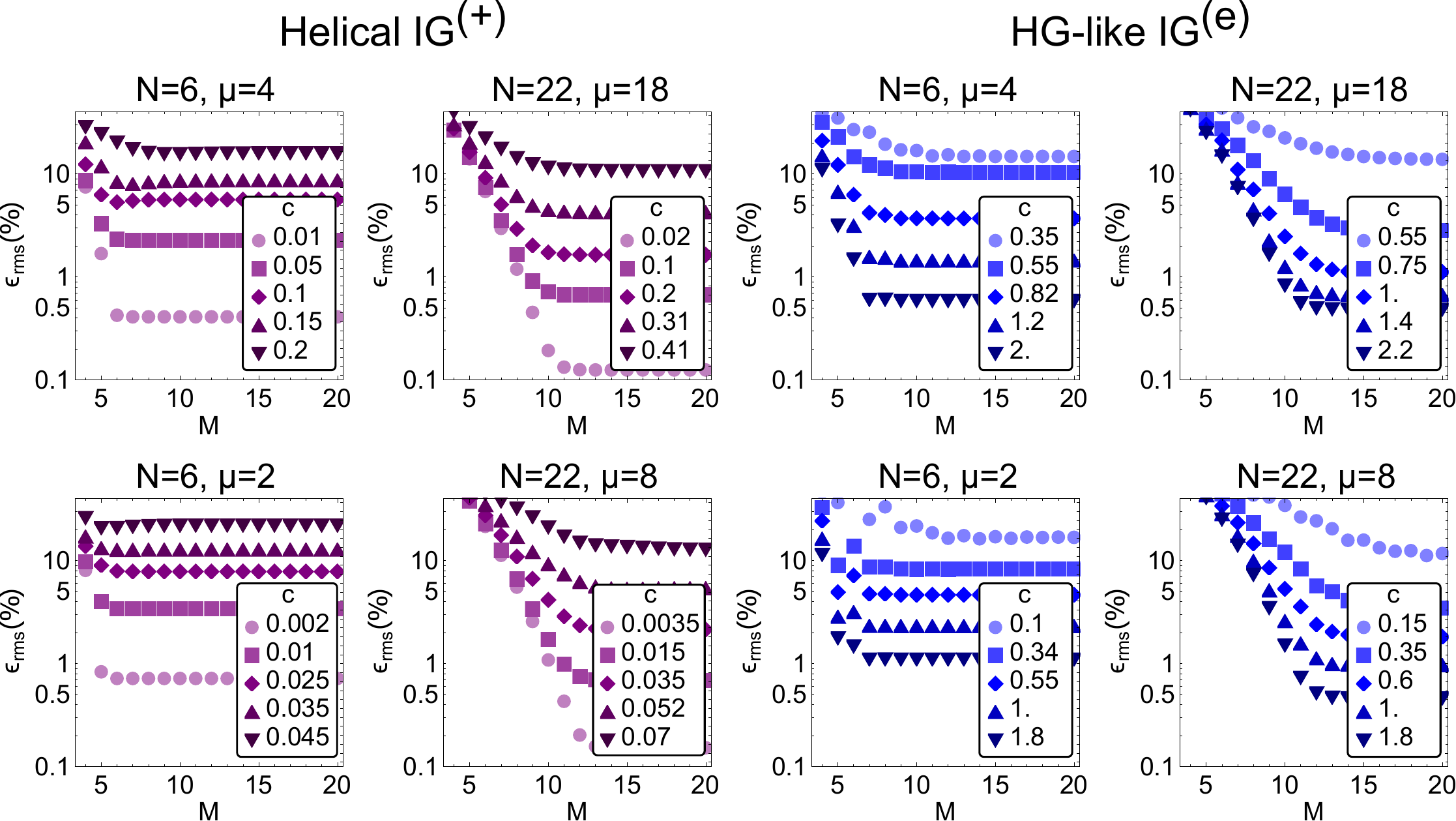}
\caption{\label{fig:error} 
Rms error $\epsilon_{\text{rms}}$ between the exact IG beams and the ray-based estimates as a function of the number of terms $M$ used
for the estimate for various values of the center $c$ for
the LG and HG-like regime and for four different IG beams.}
\end{figure}

Figure \ref{fig:rayest} also shows the excellent agreement between the ray-based estimates
and the exact fields for IG beams within the two regimes, despite the low values of $M$ being used. 
Figure \ref{fig:error} shows the rms error 
in terms of the number of contributions $M$, for beams with the same indices as in Fig.~\ref{fig:rayest} and for others with a lower value for $\mu$ for different values of $c$ (and hence of $\ve$). The plots show that in all cases the sums converge rapidly (typically for $M<N$), reaching a level of error that is due to the semiclassical approach itself, which is more accurate as the difference between $c$ and the boundary between regimes increases. This means that this approach might be computationally convenient for large $N$.

\section{Concluding remarks}

In this work we explored the structure of IG beams from the
ray-optics perspective. This ray structure embodies the main features of these beams, in particular the two separate regimes for their overall shape, as well as the boundary between them, which corresponds to a ray-optical topological transition. The analysis also reveals connections with many geometric constructions such as the Steiner and Pappus chains, Viviani curves, Keplerian orbits and medial axes, as well as  physical systems like the harmonic oscillator, the pendulum and the Bose-Hubbard dimer.  

As shown in Ref.~\cite{gutierrezcuevas2023exactly} and discussed in this work, IG beams are modes of resonant cavities whose curved mirrors contain a small amount of spherical aberration and astigmatism. Let us discuss how this fact sheds light on the structure of the beams from the ray-optical perspective. Imagine a subset of rays corresponding to a single EFR like those shown in Fig.~\ref{fig:erfs}. When confined inside an unaberrated cavity with appropriate mirror separation and curvatures, this group of rays retains its configuration after a round trip. Note though that each ray within the EFR changes position after each round trip, cycling within the EFR and exploring different values of the parameter $\tau$. In fact, except for very specific cases in which certain rational relations hold, the complete EFR can be populated by letting a single ray bounce back and forth. Now consider the effect of small amounts of astigmatism and spherical aberration in the cavity. These aberrations cause the EFR to very slightly modify its profile after each round trip. If the aberrations are in the perturbative regime, the deformed EFR will still have an elliptic profile and therefore will still constitute an EFR, one corresponding to a nearby point over the ray-Poincaré sphere. Multiple round trips then correspond to a slow displacement of the point over the ray-Poincaré sphere, which traces the PP (a generalized Viviani curve if the aberrations are astigmatism and spherical). Remarkably, the temporal rate of this slow evolution corresponds, when projected onto the $(t_1,t_2)$ plane, to the rotation or libration of a simple pendulum. That is, a ray bouncing back and forth in the aberrated cavity presents a fast cyclic evolution in $\tau$ and a slow pendulum-like precession in $\eta$. From this point of view, we see that the connection between IG modes and the pendulum does not arise in the semiclassic regime (that is, for the construction of wave modes) but in the ray-optical one.

The ray-optical picture of IG beams provides intuition not only on their behavior within a cavity but also under free propagation. For example, rays give a simple description of the transport of energy (and therefore information) over significant distances, especially for beams with sufficiently large $N$ for with the transverse profile is more structured. If one were to block a sector of a beam's profile, the change in this profile some distance away from the obstacle can be understood intuitively in terms of the rays that were blocked. Amongst other things, this explains why IG beams, like most other structured self-similar beams, present the property of ``self-healing'', in which the intensity features that were blocked reappear after some propagation distance (at the cost of others disappearing). This effect is due to the cycling under propagation of the rays within each EFR: the blocked rays associated with the missing feature are replaced by non-blocked rays, making the feature reappear. Finally, note that the ray representation of a structured beam allows the use of fast ray-based computations of the trapping forces and torques resulting from the interaction of the beam with an object whose spatial features are larger than the wavelength. 

Finally, let us remark that the description given here can be used to predict the (ray and wave) shape of self-similar beams that are modes of cavities presenting other types of aberration. By associating such aberrations with powers of the parameters $t_j$, one can easily find the shape of the PP, and from it that for the caustics by using the medial axis construction. Nevertheless, it is unlikely that any other set of aberrations would lead to beams with such a simple yet rich geometric description as the one shown here. For IG beams it is remarkable that, despite the nonlinear (that is, nonplanar) nature of the PP and its resulting different topological regimes, the caustics end up being simple confocal conic sections.

\section*{Acknowledgements}

The authors acknowledge J.M. Fellows and D.H.J. O'Dell for useful discussions. 
for useful suggestions. R.G.-C. and M.A.A. acknowledge the Excellence Initiative 
of Aix-Marseille University - A*MIDEX, a French
``Investissements d'Avenir'' programme for funding this research.
R.G.C. also acknowledges funding from the Labex WIFI 
(ANR-10-LABX-24, ANR-10-IDEX-0001-02 PSL*).

\section*{Disclosures}

The authors declare no conflicts of interest.

\section*{Data Availability Statement}

The code used to compute the exact and ray-based estimates of Ince-Gauss beams is available in Ref.~\cite{repo}.









\appendix

\section{Analytic formula for Eq.~(\ref{eq:upsi})}
\label{app:upsi}

As mentioned in the main text, it is possible to perform the integral defining the phase $\Upsilon$ analytically for the ray families describing IG fields. 
The result is given by the following expression in terms of elliptic functions:
\begin{align}
    \Upsilon(\eta) &= {\rm Re}\Bigg\{
    \tan \left[\frac{\phi(\eta)}{2}\right] \sqrt{1-c^2-R^2-2 c R \cos \phi(\eta)}+\frac{\im\, {\rm sgn}(C_+)}{\sqrt{C_+}}
   \Bigg[C_+
   E\left(\gamma
   \left|\frac{C_-}{C_+}\right.\right)\nonumber\\
   &+\frac{2 R \left(1+c^2-R^2\right) }{c-R}F\left(\gamma
   \left|\frac{C_-}{C_+}\right.\right)-\frac{4 c R }{c^2-R^2}\Pi
   \left(\frac{1-C_-}{1-C_+};\gamma
   \left|\frac{C_-}{C_+}\right.\right)\Bigg]\Bigg\},
\end{align}
where $C_\pm=1-(c\pm R)^2$, $\gamma=\im\,{\rm arcsinh}\{\tan[\phi(\eta)/2]\}$, $E$ is the elliptic integral of the second kind, $F$ elliptic integral of the first kind, and $\Pi$ is the complete integral of the third kind.
However, there is a caveat for the use of this expression: it is defined as a continuous function for $\phi$ within the interval $[-\pi,\pi]$ such that it is zero for $\phi=0$ in the LG-like regime and for $\phi=\pm\varphi_0$ in the HG-like regime. There is therefore a discontinuity at $\phi=\pm\pi$. This discontinuity poses no problem in the LG-like regime since the quantization of the solid angle ensures that, when placed in the exponent of the semiclassical estimate, the resulting phase discontinuity is an integer multiple of $2\pi$. For the HG-like regime, on the other hand, the discontinuity must be removed by hand to make the function continuous at $\phi=\pm\pi$. Unfortunately the limit $\phi\to\pm\pi$ in this analytic form is complicated to evaluate, so the constant that must be added/subtracted must be evaluated numerically by using a value very near this limit.

\section{Derivation of the amplitude equation}
\label{app:amp}

To compute the ray-based estimate, we employ a Gaussian summation method where each ray is dressed  with a Gaussian contribution of the form
\begin{align}
g (\bt r ; \bt Q, \bt P) =  
\sqrt{ \frac{k \Gamma}{2 \pi}} \exp \left[ - \frac{k \Gamma }{2}\norm{\bt r - \bt Q}^2 + \im k (\bt r - \bt Q) \cdot \bt P \right] e^{\im k L(\eta , \tau)}. 
\end{align}
The center $\bt Q$  and direction $\bt P$ of the Gaussians are the same as those of the corresponding rays. $\Gamma$ is a free parameter that can be chosen to optimize the field estimate. For the study of SG beams, its optimal value is the inverse Rayleigh range, $\Gamma = 2/k w_0^2$. Given this choice, these Gaussian contributions correspond to displaced coherent states of the two-dimensional isotropic harmonic oscillator. As opposed to what is done in the main text, here we keep the dependence on both parameters of the ray family so that the estimate is given by
\begin{align}
U(\bt r) =  \iint A(\eta) \sqrt{J(\tau,\eta)}  \;g (\bt r ; \bt Q, \bt P) \ud \tau \ud \eta 
\end{align}
where $L(\eta , \tau)$ is the optical path length that plays the role of the action in mechanics. The weight $A$ only depends on the parameter $\eta$ so that all the rays in the same ellipse are weighted equally, which is a necessary condition to obtain a self-similar beam.

To determine the appropriate equation for the amplitude, we substitute the semiclassical estimate into the Ince equation,
\begin{align}
\iint A(\eta) \sqrt{J(\tau,\eta)}  \bigg[ \hat T_3^2  g (\bt r ; \bt Q, \bt P) +\frac{\varepsilon}{2}\hat T_1  g (\bt r ; \bt Q, \bt P) -\frac{a}{4} g (\bt r ; \bt Q, \bt P)\bigg] \ud \tau \ud \eta =0 .
\end{align}
Since $g$ is the only part that has a spatial dependence, we start by computing the following relations:
\bse
\label{eq:ts2g}
\begin{align}
 \hat T_1 g(\bt r, \eta, \tau) 
=& 
\left[ \frac{N+1}{2}	 ( |
z_x|^2   -|z_y|^2 )+ \frac{\sqrt{N+1}}{w} \left( z_x \Delta_x - z_y \Delta_y \right) \right]g(\bt r, \eta, \tau)   \\
 \hat T_3^2 g(\bs r, \eta, \tau) 
=& \bigg[ (N+1)^{2}(p_xq_y-p_yq_x)^2  +\frac{ N+1}{2} \bt q \cdot \bt z  
-\frac{2\im(N+1)^{3/2}}{w} (p_xq_y-p_yq_x)(z_x \Delta_y -z_y \Delta_x) \nonumber
\\
&
 +\frac{N+1}{w^2}(2z_xz_y \Delta_x\Delta_y-z_x^{2}\Delta_y^2-z_y^{2}\Delta_x^2  )
 +\frac{\sqrt{N+1}}{2w}  \bt z \cdot \bs \Delta
\bigg]g(\bt r, \eta, \tau) 
\end{align}
\ese
where we introduced $\bs \Delta=(\Delta_x , \Delta_y)= \bt r -\bt Q$, 
\begin{align}
\bt Z= \Gamma \bt Q +\im \bt P =\frac{2}{wk} \sqrt{N+1} \bt v (\eta) e^{-\im \tau}=\frac{2}{wk} \sqrt{N+1} ( \bt q +\im \bt p)=\frac{2}{wk} \sqrt{N+1} \bt z
\end{align}
 and wrote explicitly the dependence of $g$ on $\eta$ and $\tau$. 
 
The next step is to move all the spatial dependence to $g$, in order to be able to identify part of the integrand as being equal to zero. This is done by using the following identities,
\begin{align}
 \Delta_x g=
  \frac{w}{2(N+1)^{1/2}}  \bigg( \frac{iz_y}{j}\pd{g}{\eta} +\frac{z'_y}{j}\pd{g}{\tau} \bigg),
 \qquad
\Delta_y g=
 - \frac{w}{2(N+1)^{1/2}}  \bigg(\frac{iz_x}{j}\pd{g}{\eta}+\frac{z'_x}{j} \pd{g}{\tau} \bigg)
\end{align}
where primes indicate derivatives with respect to $\eta$ (all derivatives with respect to $\tau$ can be taken explicitly because the dependence of $\bt Q$ and $\bt P$ on  $\tau$ is  known) and
\begin{align}
J(\tau,\eta) =\pd{\bs Z}{(\tau,\eta)}=\pd{Z_x}{\tau} \pd{Z_y}{\eta}-\pd{Z_y}{\tau}\pd{Z_x}{\eta}
=\frac{4 \im (N+1)}{w^2k^2} \left(z_y\pd{z_x}{\eta}- z_x \pd{z_y}{\eta}\right)
=\frac{4 (N+1)}{w^2k^2} j(\tau ,\eta ) ,
\end{align}
and then integrating by parts using the fact that the integrations are periodic. This leads to the following identities which apply to all the terms appearing in Eq.~(\ref{eq:ts2g})
\begin{align*}
\iint  A F \sqrt{J}g \,\mathrm {d} \eta \mathrm {d} \tau
=& \frac{2 (N+1)^{1/2}}{wk} \iint  A F \sqrt{j}g \,\mathrm {d} \eta \mathrm {d} \tau
\\
\iint  A F \sqrt{J}\Delta_x g \,\mathrm {d} \eta \mathrm {d} \tau
=&-\frac{1}{k}\iint \left[ iF'z_y\frac{A}{\sqrt{j } }  + iFz_y\left(\frac{A}{\sqrt{j } } \right)'  \right]g  \,\mathrm {d} \eta \mathrm {d} \tau
\\
\iint  A F \sqrt{J}\Delta_y g \,\mathrm {d} \eta \mathrm {d} \tau
=&\frac{1}{k}\iint \left[ iF'z_x\frac{A}{\sqrt{j } } + iFz_x\left(\frac{A}{\sqrt{j } } \right)' \right]g  \,\mathrm {d} \eta \mathrm {d} \tau
\end{align*}
where $F(\eta,\tau) =f(\eta)\exp(-\im \tau)$, with $f$ being an arbitrary function of $\eta$, and for the rest of the terms we can be more specific,
\begin{align*}
\iint A \sqrt{J } \bt z \cdot \bs \Delta g \,\mathrm {d} \eta \mathrm {d} \tau
=&-\frac{1}{k}\iint j \frac{A}{\sqrt{j }} g  \,\mathrm {d} \eta \mathrm {d} \tau \\
\iint  Az_y^2 \sqrt{J}\Delta_x^2 g \,\mathrm {d} \eta \mathrm {d} \tau=& \frac{w(N+1)^{1/2}}{k}\iint \Bigg\{z_y^2 \frac{A}{\sqrt{j } } \left[ iq_x'z_y+ p_x z'_y \right]  \Bigg\} g\,\mathrm {d} \eta \mathrm {d} \tau \\
&- \frac{w}{2 k (N+1)^{1/2}}\iint  \Bigg\{
\frac{z_y}{3}\left[ \frac{(z_y^3)'}{j}\right]' \left(\frac{A}{\sqrt{j } }\right)+\left( \frac{z_y^4}{j}\right)' \left(\frac{A}{\sqrt{j } }\right)' \\ 
& +\left( \frac{z_y^4}{j}\right) \left(\frac{A}{\sqrt{j} }\right)'' \Bigg\} g
   \,\mathrm {d} \eta \mathrm {d} \tau
\\
\iint    Az_x^2 \sqrt{J}\Delta_y^2 g \,\mathrm {d} \eta \mathrm {d} \tau
=& -\frac{w(N+1)^{1/2}}{k}\iint  \Bigg\{z_x^2 \frac{A}{\sqrt{j } } \left[ iq_y'z_x+ p_y z'_x \right]  \Bigg\} g\,\mathrm {d} \eta \mathrm {d} \tau \\
&-  \frac{w}{2 k (N+1)^{1/2}}\iint \Bigg\{
\frac{z_x}{3}\left[ \frac{(z_x^3)'}{j}\right]' \left(\frac{A}{\sqrt{j } }\right)+\left( \frac{z_x^4}{J}\right)' \left(\frac{A}{\sqrt{j } }\right)' \\
& +\left( \frac{z_x^4}{j}\right) \left(\frac{A}{\sqrt{j } }\right)'' \Bigg\} g
   \,\mathrm {d} \eta \mathrm {d} \tau
\end{align*}
\begin{align*}
\iint   Az_x z_y \sqrt{J}\Delta_x \Delta_y g \,\mathrm {d} \eta \mathrm {d} \tau
=&\frac{w(N+1)^{1/2}}{k}\iint  \frac{i}{2} z_x z_y (|z_y|^2)' \frac{A}{\sqrt{j } }g  \,\mathrm {d} \eta \mathrm {d} \tau\\
& +\frac{w}{2 k (N+1)^{1/2}}\iint   \Bigg\{ \frac{z_xz_y}{2j^2}\bigg\{j(z_x''z_y+z_xz_y'')+4jz_x'z_y'-j'(z_xz_y)'\bigg\} \frac{A}{\sqrt{j } }\\
&+  \left(\frac{z_x^2 z_y^2}{j}\right) '  \left(\frac{A}{\sqrt{j } } \right)'  + \left(\frac{z_x^2 z_y^2}{j}\right)  \left(\frac{A}{\sqrt{j } } \right)'' \Bigg\}g 
 \,\mathrm {d} \eta \mathrm {d} \tau.
\end{align*}

Using the previously derived identities, all the spatial dependence can be passed onto $g$ so that it can be factored out along  with $A$ and $j^{1/2}$. The rest of the integrand can then be set equal to zero. Moreover, since the classical 
(or ray) limit is attained for large $N$, the contributions of different orders can be treated separately. 
The first order leads to the following equation,
\begin{align}
(N+1)^{2}(p_xq_y-p_yq_x)^2 +\frac{\varepsilon}{2} \frac{N+1}{2}	 ( |z_x|^2   -|z_y|^2 ) = \frac{a}{4}  .
\end{align}
Noting that
\begin{align}
T_3^2=(N+1)^2 (p_xq_y-p_yq_x)^2 \qquad T_1 =\frac{N+1}{2} (|z_x|^2-|z_y|^2)
\end{align}
it is clear that the previous equation is nothing more than the PP path.
The following term requires  more work, but after some tedious mathematical manipulations we get
\begin{align*}
0
=& \Bigg\{  \Big[  (p_xq_y-p_yq_x)( z_x^2+z_y^2)   \Big]'  \frac{A}{\sqrt{j } }
+ 2 \bigg[ (p_xq_y-p_yq_x) (z_x^2+ z_y^2 ) \bigg]  \left(\frac{A}{\sqrt{j } } \right)' 
 \Bigg\} \\
&- \frac{\varepsilon \im}{2(N+1)} \left[   (z_xz_y)'\frac{A}{\sqrt{j } }  + 2 z_xz_y\left(\frac{A}{\sqrt{j } } \right)'  
  \right].
\end{align*}
This equation says how the weight function changes depending on the parametrization of the PP. This equation can be integrated after we multiply it by $A/j^{1/2}$, thus leading to
\begin{align*}
 A^2
=& \frac{B j}{  (p_xq_y-p_yq_x)( z_x^2+z_y^2)  - \frac{\varepsilon \im}{2(N+1)} z_xz_y },
\end{align*}
with $B$ being a constant of integration, and for the remainder of the derivation we will set $B=\im$ without loss of generality. 

Both the equation for the PP and that for the amplitude $A$ depend only on $\eta$, as can be appreciated when writing them in terms of  $\theta$ and $\phi$,
\bse
\begin{align}
(N+1)^{2}\cos^2 \theta +{\varepsilon} (N+1)	\cos \phi \sin \theta = a  \\
 A^2
= \frac{- \sin \theta \phi' + \im\theta'}{  \cos \theta \sin \theta  + \frac{\varepsilon \im}{2(N+1)}(\im \cos \phi \cos \theta + \sin \phi)}.
\end{align}
\ese
Taking the derivative of the first equation
\begin{align}
 \cos \theta \sin \theta =\frac{\varepsilon}{2 (N+1)} 	( -\frac{\phi'}{\theta'} 	\sin \phi \sin \theta +\cos \phi \cos \theta)  
\end{align}
and substituting it into the second one we get
\begin{align}
 A^2
=\frac{2  (N+1) } {\varepsilon  }
 \frac{\theta'  }{  \sin \phi }.
\end{align}
Finally, by writing this equation in terms of $\varphi$, we have that
\begin{align}
 A^2
= \pm 
 \frac{\varphi'  }{  \sqrt{ 1 - c^2 -R^2 - 2 c R \cos \varphi } },
\end{align}
where we used the relation $c=\varepsilon/2(N+1)$. 
The denominator is proportional to the angular velocity of a simple pendulum, therefore if we set $\varphi= \eta$ then the amplitude of the ray-ellipses is inversely proportional to the square root of the angular velocity of the pendulum. Another way to look at it is to assume that the amplitude is constant, say $A=1$, then the solution of the angular variable $\varphi$ follows the dynamics of the pendulum where $\eta$ plays the role of time. The connection to the simple pendulum becomes clearer after taking the second derivative of the previous equation, leading to 
\begin{align}
\varphi''-  c R  \sin \varphi = 0.
\end{align}
This is the defining equation of a simple pendulum.

\bibliography{Refs-raysNwaves}

\begin{thebibliography}{63}%
\makeatletter
\providecommand \@ifxundefined [1]{%
 \@ifx{#1\undefined}
}%
\providecommand \@ifnum [1]{%
 \ifnum #1\expandafter \@firstoftwo
 \else \expandafter \@secondoftwo
 \fi
}%
\providecommand \@ifx [1]{%
 \ifx #1\expandafter \@firstoftwo
 \else \expandafter \@secondoftwo
 \fi
}%
\providecommand \natexlab [1]{#1}%
\providecommand \enquote  [1]{``#1''}%
\providecommand \bibnamefont  [1]{#1}%
\providecommand \bibfnamefont [1]{#1}%
\providecommand \citenamefont [1]{#1}%
\providecommand \href@noop [0]{\@secondoftwo}%
\providecommand \href [0]{\begingroup \@sanitize@url \@href}%
\providecommand \@href[1]{\@@startlink{#1}\@@href}%
\providecommand \@@href[1]{\endgroup#1\@@endlink}%
\providecommand \@sanitize@url [0]{\catcode `\\12\catcode `\$12\catcode
  `\&12\catcode `\#12\catcode `\^12\catcode `\_12\catcode `\%12\relax}%
\providecommand \@@startlink[1]{}%
\providecommand \@@endlink[0]{}%
\providecommand \url  [0]{\begingroup\@sanitize@url \@url }%
\providecommand \@url [1]{\endgroup\@href {#1}{\urlprefix }}%
\providecommand \urlprefix  [0]{URL }%
\providecommand \Eprint [0]{\href }%
\providecommand \doibase [0]{https://doi.org/}%
\providecommand \selectlanguage [0]{\@gobble}%
\providecommand \bibinfo  [0]{\@secondoftwo}%
\providecommand \bibfield  [0]{\@secondoftwo}%
\providecommand \translation [1]{[#1]}%
\providecommand \BibitemOpen [0]{}%
\providecommand \bibitemStop [0]{}%
\providecommand \bibitemNoStop [0]{.\EOS\space}%
\providecommand \EOS [0]{\spacefactor3000\relax}%
\providecommand \BibitemShut  [1]{\csname bibitem#1\endcsname}%
\let\auto@bib@innerbib\@empty
\bibitem [{\citenamefont {Boyer}\ \emph
  {et~al.}(1975{\natexlab{a}})\citenamefont {Boyer}, \citenamefont {Kalnins},\
  and\ \citenamefont {Miller}}]{boyer1975liea}%
  \BibitemOpen
  \bibfield  {author} {\bibinfo {author} {\bibfnamefont {C.~P.}\ \bibnamefont
  {Boyer}}, \bibinfo {author} {\bibfnamefont {E.~G.}\ \bibnamefont {Kalnins}},\
  and\ \bibinfo {author} {\bibfnamefont {W.}~\bibnamefont {Miller}},\
  }\bibfield  {title} {\bibinfo {title} {Lie theory and separation of
  variables. 6. the equation $i u_t + \delta_2 u = 0$},\ }\href
  {https://doi.org/10.1063/1.522573} {\bibfield  {journal} {\bibinfo  {journal}
  {J. Math. Phys.}\ }\textbf {\bibinfo {volume} {16}},\ \bibinfo {pages} {499}
  (\bibinfo {year} {1975}{\natexlab{a}})}\BibitemShut {NoStop}%
\bibitem [{\citenamefont {Siegman}(1986)}]{siegman1986lasers}%
  \BibitemOpen
  \bibfield  {author} {\bibinfo {author} {\bibfnamefont {A.~E.}\ \bibnamefont
  {Siegman}},\ }\href@noop {} {\emph {\bibinfo {title} {Lasers}}}\ (\bibinfo
  {publisher} {University Science Books},\ \bibinfo {address} {Sausalito, CA},\
  \bibinfo {year} {1986})\BibitemShut {NoStop}%
\bibitem [{\citenamefont {Siviloglou}\ \emph {et~al.}(2007)\citenamefont
  {Siviloglou}, \citenamefont {Broky}, \citenamefont {Dogariu},\ and\
  \citenamefont {Christodoulides}}]{siviloglou2007observation}%
  \BibitemOpen
  \bibfield  {author} {\bibinfo {author} {\bibfnamefont {G.~A.}\ \bibnamefont
  {Siviloglou}}, \bibinfo {author} {\bibfnamefont {J.}~\bibnamefont {Broky}},
  \bibinfo {author} {\bibfnamefont {A.}~\bibnamefont {Dogariu}},\ and\ \bibinfo
  {author} {\bibfnamefont {D.~N.}\ \bibnamefont {Christodoulides}},\ }\bibfield
   {title} {\bibinfo {title} {Observation of accelerating airy beams},\ }\href
  {https://doi.org/10.1103/physrevlett.99.213901} {\bibfield  {journal}
  {\bibinfo  {journal} {Phys. Rev. Lett.}\ }\textbf {\bibinfo {volume} {99}},\
  \bibinfo {pages} {213901} (\bibinfo {year} {2007})}\BibitemShut {NoStop}%
\bibitem [{\citenamefont {Bandres}(2008)}]{bandres2008accelerating}%
  \BibitemOpen
  \bibfield  {author} {\bibinfo {author} {\bibfnamefont {M.~A.}\ \bibnamefont
  {Bandres}},\ }\bibfield  {title} {\bibinfo {title} {Accelerating parabolic
  beams},\ }\href {https://doi.org/10.1364/ol.33.001678} {\bibfield  {journal}
  {\bibinfo  {journal} {Opt. Lett.}\ }\textbf {\bibinfo {volume} {33}},\
  \bibinfo {pages} {1678} (\bibinfo {year} {2008})}\BibitemShut {NoStop}%
\bibitem [{\citenamefont {Levy}\ \emph {et~al.}(2016)\citenamefont {Levy},
  \citenamefont {Derevyanko},\ and\ \citenamefont
  {Silberberg}}]{levy2016light}%
  \BibitemOpen
  \bibfield  {author} {\bibinfo {author} {\bibfnamefont {U.}~\bibnamefont
  {Levy}}, \bibinfo {author} {\bibfnamefont {S.}~\bibnamefont {Derevyanko}},\
  and\ \bibinfo {author} {\bibfnamefont {Y.}~\bibnamefont {Silberberg}},\
  }\bibfield  {title} {\bibinfo {title} {Light modes of free space},\ }in\
  \href {https://doi.org/10.1016/bs.po.2015.10.001} {\emph {\bibinfo
  {booktitle} {Prog. Opt.}}}\ (\bibinfo  {publisher} {Elsevier},\ \bibinfo
  {year} {2016})\ pp.\ \bibinfo {pages} {237--281}\BibitemShut {NoStop}%
\bibitem [{\citenamefont {Dennis}\ and\ \citenamefont
  {Ring}(2013)}]{dennis2013propagation}%
  \BibitemOpen
  \bibfield  {author} {\bibinfo {author} {\bibfnamefont {M.~R.}\ \bibnamefont
  {Dennis}}\ and\ \bibinfo {author} {\bibfnamefont {J.~D.}\ \bibnamefont
  {Ring}},\ }\bibfield  {title} {\bibinfo {title} {Propagation-invariant beams
  with quantum pendulum spectra: from bessel beams to gaussian beam-beams},\
  }\href {https://doi.org/10.1364/ol.38.003325} {\bibfield  {journal} {\bibinfo
   {journal} {Opt. Lett.}\ }\textbf {\bibinfo {volume} {38}},\ \bibinfo {pages}
  {3325} (\bibinfo {year} {2013})}\BibitemShut {NoStop}%
\bibitem [{\citenamefont {Guti{\'{e}}rrez-Cuevas}\ and\ \citenamefont
  {Alonso}(2017{\natexlab{a}})}]{gutierrez-cuevas2017polynomials}%
  \BibitemOpen
  \bibfield  {author} {\bibinfo {author} {\bibfnamefont {R.}~\bibnamefont
  {Guti{\'{e}}rrez-Cuevas}}\ and\ \bibinfo {author} {\bibfnamefont {M.~A.}\
  \bibnamefont {Alonso}},\ }\bibfield  {title} {\bibinfo {title} {Polynomials
  of {G}aussians and vortex-{G}aussian beams as complete, transversely confined
  bases},\ }\href {https://doi.org/10.1364/ol.42.002205} {\bibfield  {journal}
  {\bibinfo  {journal} {Opt. Lett.}\ }\textbf {\bibinfo {volume} {42}},\
  \bibinfo {pages} {2205} (\bibinfo {year} {2017}{\natexlab{a}})}\BibitemShut
  {NoStop}%
\bibitem [{\citenamefont {Guti{\'{e}}rrez-Cuevas}\ and\ \citenamefont
  {Alonso}(2017{\natexlab{b}})}]{gutierrez-cuevas2017complete}%
  \BibitemOpen
  \bibfield  {author} {\bibinfo {author} {\bibfnamefont {R.}~\bibnamefont
  {Guti{\'{e}}rrez-Cuevas}}\ and\ \bibinfo {author} {\bibfnamefont {M.~A.}\
  \bibnamefont {Alonso}},\ }\bibfield  {title} {\bibinfo {title} {Complete
  confined bases for beam propagation in {C}artesian coordinates},\ }\href
  {https://doi.org/10.1364/josaa.34.001697} {\bibfield  {journal} {\bibinfo
  {journal} {J. Opt. Soc. Am. A}\ }\textbf {\bibinfo {volume} {34}},\ \bibinfo
  {pages} {1697} (\bibinfo {year} {2017}{\natexlab{b}})}\BibitemShut {NoStop}%
\bibitem [{\citenamefont {Boyer}\ \emph
  {et~al.}(1975{\natexlab{b}})\citenamefont {Boyer}, \citenamefont {Kalnins},\
  and\ \citenamefont {Miller}}]{boyer1975lie}%
  \BibitemOpen
  \bibfield  {author} {\bibinfo {author} {\bibfnamefont {C.~P.}\ \bibnamefont
  {Boyer}}, \bibinfo {author} {\bibfnamefont {E.~G.}\ \bibnamefont {Kalnins}},\
  and\ \bibinfo {author} {\bibfnamefont {W.}~\bibnamefont {Miller}},\
  }\bibfield  {title} {\bibinfo {title} {Lie theory and separation of
  variables. 7. the harmonic oscillator in elliptic coordinates and ince
  polynomials},\ }\href {https://doi.org/10.1063/1.522574} {\bibfield
  {journal} {\bibinfo  {journal} {J. Math. Phys.}\ }\textbf {\bibinfo {volume}
  {16}},\ \bibinfo {pages} {512} (\bibinfo {year}
  {1975}{\natexlab{b}})}\BibitemShut {NoStop}%
\bibitem [{\citenamefont {Andrews}(2008)}]{andrews2008structured}%
  \BibitemOpen
  \bibfield  {author} {\bibinfo {author} {\bibfnamefont {D.~L.}\ \bibnamefont
  {Andrews}},\ }\href
  {https://www.ebook.de/de/product/7336177/david_l_andrews_structured_light_and_its_applications_an_introduction_to_phase_structured_beams_and_nanoscale_optical_forces.html}
  {\emph {\bibinfo {title} {Structured Light and Its Applications: An
  Introduction to Phase-Structured Beams and Nanoscale Optical Forces}}}\
  (\bibinfo  {publisher} {ACADEMIC PR INC},\ \bibinfo {year}
  {2008})\BibitemShut {NoStop}%
\bibitem [{\citenamefont {Berkhout}\ \emph {et~al.}(2010)\citenamefont
  {Berkhout}, \citenamefont {Lavery}, \citenamefont {Courtial}, \citenamefont
  {Beijersbergen},\ and\ \citenamefont {Padgett}}]{berkhout2010efficient}%
  \BibitemOpen
  \bibfield  {author} {\bibinfo {author} {\bibfnamefont {G.~C.~G.}\
  \bibnamefont {Berkhout}}, \bibinfo {author} {\bibfnamefont {M.~P.~J.}\
  \bibnamefont {Lavery}}, \bibinfo {author} {\bibfnamefont {J.}~\bibnamefont
  {Courtial}}, \bibinfo {author} {\bibfnamefont {M.~W.}\ \bibnamefont
  {Beijersbergen}},\ and\ \bibinfo {author} {\bibfnamefont {M.~J.}\
  \bibnamefont {Padgett}},\ }\bibfield  {title} {\bibinfo {title} {Efficient
  sorting of orbital angular momentum states of light},\ }\href
  {https://doi.org/10.1103/physrevlett.105.153601} {\bibfield  {journal}
  {\bibinfo  {journal} {Phys. Rev. Lett.}\ }\textbf {\bibinfo {volume} {105}},\
  \bibinfo {pages} {153601} (\bibinfo {year} {2010})}\BibitemShut {NoStop}%
\bibitem [{\citenamefont {Zhou}\ \emph {et~al.}(2017)\citenamefont {Zhou},
  \citenamefont {Mirhosseini}, \citenamefont {Fu}, \citenamefont {Zhao},
  \citenamefont {Rafsanjani}, \citenamefont {Willner},\ and\ \citenamefont
  {Boyd}}]{zhou2017sorting}%
  \BibitemOpen
  \bibfield  {author} {\bibinfo {author} {\bibfnamefont {Y.}~\bibnamefont
  {Zhou}}, \bibinfo {author} {\bibfnamefont {M.}~\bibnamefont {Mirhosseini}},
  \bibinfo {author} {\bibfnamefont {D.}~\bibnamefont {Fu}}, \bibinfo {author}
  {\bibfnamefont {J.}~\bibnamefont {Zhao}}, \bibinfo {author} {\bibfnamefont
  {S.~M.~H.}\ \bibnamefont {Rafsanjani}}, \bibinfo {author} {\bibfnamefont
  {A.~E.}\ \bibnamefont {Willner}},\ and\ \bibinfo {author} {\bibfnamefont
  {R.~W.}\ \bibnamefont {Boyd}},\ }\bibfield  {title} {\bibinfo {title}
  {Sorting photons by radial quantum number},\ }\href
  {https://doi.org/10.1103/physrevlett.119.263602} {\bibfield  {journal}
  {\bibinfo  {journal} {Phys. Rev. Lett.}\ }\textbf {\bibinfo {volume} {119}},\
  \bibinfo {pages} {263602} (\bibinfo {year} {2017})}\BibitemShut {NoStop}%
\bibitem [{\citenamefont {Gu}\ \emph {et~al.}(2018)\citenamefont {Gu},
  \citenamefont {Krenn}, \citenamefont {Erhard},\ and\ \citenamefont
  {Zeilinger}}]{gu2018gouy}%
  \BibitemOpen
  \bibfield  {author} {\bibinfo {author} {\bibfnamefont {X.}~\bibnamefont
  {Gu}}, \bibinfo {author} {\bibfnamefont {M.}~\bibnamefont {Krenn}}, \bibinfo
  {author} {\bibfnamefont {M.}~\bibnamefont {Erhard}},\ and\ \bibinfo {author}
  {\bibfnamefont {A.}~\bibnamefont {Zeilinger}},\ }\bibfield  {title} {\bibinfo
  {title} {Gouy phase radial mode sorter for light: Concepts and experiments},\
  }\href {https://doi.org/10.1103/physrevlett.120.103601} {\bibfield  {journal}
  {\bibinfo  {journal} {Phys. Rev. Lett.}\ }\textbf {\bibinfo {volume} {120}},\
  \bibinfo {pages} {103601} (\bibinfo {year} {2018})}\BibitemShut {NoStop}%
\bibitem [{\citenamefont {Tsang}\ \emph {et~al.}(2016)\citenamefont {Tsang},
  \citenamefont {Nair},\ and\ \citenamefont {Lu}}]{tsang2016quantum}%
  \BibitemOpen
  \bibfield  {author} {\bibinfo {author} {\bibfnamefont {M.}~\bibnamefont
  {Tsang}}, \bibinfo {author} {\bibfnamefont {R.}~\bibnamefont {Nair}},\ and\
  \bibinfo {author} {\bibfnamefont {X.-M.}\ \bibnamefont {Lu}},\ }\bibfield
  {title} {\bibinfo {title} {Quantum theory of superresolution for two
  incoherent optical point sources},\ }\href
  {https://doi.org/10.1103/physrevx.6.031033} {\bibfield  {journal} {\bibinfo
  {journal} {Phys. Rev. X}\ }\textbf {\bibinfo {volume} {6}},\ \bibinfo {pages}
  {031033} (\bibinfo {year} {2016})}\BibitemShut {NoStop}%
\bibitem [{\citenamefont {Yao}\ and\ \citenamefont
  {Padgett}(2011)}]{yao2011orbital}%
  \BibitemOpen
  \bibfield  {author} {\bibinfo {author} {\bibfnamefont {A.~M.}\ \bibnamefont
  {Yao}}\ and\ \bibinfo {author} {\bibfnamefont {M.~J.}\ \bibnamefont
  {Padgett}},\ }\bibfield  {title} {\bibinfo {title} {Orbital angular momentum:
  origins, behavior and applications},\ }\href
  {https://doi.org/10.1364/aop.3.000161} {\bibfield  {journal} {\bibinfo
  {journal} {Adv. Opt. Photonics}\ }\textbf {\bibinfo {volume} {3}},\ \bibinfo
  {pages} {161} (\bibinfo {year} {2011})}\BibitemShut {NoStop}%
\bibitem [{\citenamefont {Berry}\ and\ \citenamefont
  {McDonald}(2008)}]{berry2008exact}%
  \BibitemOpen
  \bibfield  {author} {\bibinfo {author} {\bibfnamefont {M.~V.}\ \bibnamefont
  {Berry}}\ and\ \bibinfo {author} {\bibfnamefont {K.~T.}\ \bibnamefont
  {McDonald}},\ }\bibfield  {title} {\bibinfo {title} {Exact and geometrical
  optics energy trajectories in twisted beams},\ }\href
  {https://doi.org/10.1088/1464-4258/10/3/035005} {\bibfield  {journal}
  {\bibinfo  {journal} {J. Opt.}\ }\textbf {\bibinfo {volume} {10}},\ \bibinfo
  {pages} {035005} (\bibinfo {year} {2008})}\BibitemShut {NoStop}%
\bibitem [{\citenamefont {Alonso}\ and\ \citenamefont
  {Dennis}(2017)}]{alonso2017ray}%
  \BibitemOpen
  \bibfield  {author} {\bibinfo {author} {\bibfnamefont {M.~A.}\ \bibnamefont
  {Alonso}}\ and\ \bibinfo {author} {\bibfnamefont {M.~R.}\ \bibnamefont
  {Dennis}},\ }\bibfield  {title} {\bibinfo {title} {Ray-optical
  {P}oincar{\'{e}} sphere for structured {G}aussian beams},\ }\href
  {https://doi.org/10.1364/optica.4.000476} {\bibfield  {journal} {\bibinfo
  {journal} {Optica}\ }\textbf {\bibinfo {volume} {4}},\ \bibinfo {pages} {476}
  (\bibinfo {year} {2017})}\BibitemShut {NoStop}%
\bibitem [{\citenamefont {Dennis}\ and\ \citenamefont
  {Alonso}(2019)}]{dennis2019gaussian}%
  \BibitemOpen
  \bibfield  {author} {\bibinfo {author} {\bibfnamefont {M.~R.}\ \bibnamefont
  {Dennis}}\ and\ \bibinfo {author} {\bibfnamefont {M.~A.}\ \bibnamefont
  {Alonso}},\ }\bibfield  {title} {\bibinfo {title} {Gaussian mode families
  from systems of rays},\ }\href {https://doi.org/10.1088/2515-7647/ab011d}
  {\bibfield  {journal} {\bibinfo  {journal} {J. Phys. Photonics}\ }\textbf
  {\bibinfo {volume} {1}},\ \bibinfo {pages} {025003} (\bibinfo {year}
  {2019})}\BibitemShut {NoStop}%
\bibitem [{\citenamefont {Arscott}(1964)}]{arscott1964periodic}%
  \BibitemOpen
  \bibfield  {author} {\bibinfo {author} {\bibfnamefont {F.~M.}\ \bibnamefont
  {Arscott}},\ }\href@noop {} {\emph {\bibinfo {title} {Periodic Differential
  Equations}}}\ (\bibinfo  {publisher} {Pergamon Press},\ \bibinfo {year}
  {1964})\BibitemShut {NoStop}%
\bibitem [{\citenamefont {Bandres}\ and\ \citenamefont
  {Guti{\'{e}}rrez-Vega}(2004{\natexlab{a}})}]{bandres2004incegaussiana}%
  \BibitemOpen
  \bibfield  {author} {\bibinfo {author} {\bibfnamefont {M.~A.}\ \bibnamefont
  {Bandres}}\ and\ \bibinfo {author} {\bibfnamefont {J.~C.}\ \bibnamefont
  {Guti{\'{e}}rrez-Vega}},\ }\bibfield  {title} {\bibinfo {title}
  {Ince{\textendash}{G}aussian beams},\ }\href
  {https://doi.org/10.1364/ol.29.000144} {\bibfield  {journal} {\bibinfo
  {journal} {Opt. Lett.}\ }\textbf {\bibinfo {volume} {29}},\ \bibinfo {pages}
  {144} (\bibinfo {year} {2004}{\natexlab{a}})}\BibitemShut {NoStop}%
\bibitem [{\citenamefont {Bandres}\ and\ \citenamefont
  {Guti{\'{e}}rrez-Vega}(2004{\natexlab{b}})}]{bandres2004incegaussian}%
  \BibitemOpen
  \bibfield  {author} {\bibinfo {author} {\bibfnamefont {M.~A.}\ \bibnamefont
  {Bandres}}\ and\ \bibinfo {author} {\bibfnamefont {J.~C.}\ \bibnamefont
  {Guti{\'{e}}rrez-Vega}},\ }\bibfield  {title} {\bibinfo {title}
  {Ince{\textendash}{G}aussian modes of the paraxial wave equation and stable
  resonators},\ }\href {https://doi.org/10.1364/josaa.21.000873} {\bibfield
  {journal} {\bibinfo  {journal} {J. Opt. Soc. Am. A}\ }\textbf {\bibinfo
  {volume} {21}},\ \bibinfo {pages} {873} (\bibinfo {year}
  {2004}{\natexlab{b}})}\BibitemShut {NoStop}%
\bibitem [{\citenamefont {Yao-Li}\ \emph {et~al.}(2020)\citenamefont {Yao-Li},
  \citenamefont {Hu}, \citenamefont {Perez-Garcia}, \citenamefont {Bo-Zhao},
  \citenamefont {Gao}, \citenamefont {Zhu},\ and\ \citenamefont
  {Rosales-Guzm{\'{a}}n}}]{yaoli2020classically}%
  \BibitemOpen
  \bibfield  {author} {\bibinfo {author} {\bibnamefont {Yao-Li}}, \bibinfo
  {author} {\bibfnamefont {X.-B.}\ \bibnamefont {Hu}}, \bibinfo {author}
  {\bibfnamefont {B.}~\bibnamefont {Perez-Garcia}}, \bibinfo {author}
  {\bibnamefont {Bo-Zhao}}, \bibinfo {author} {\bibfnamefont {W.}~\bibnamefont
  {Gao}}, \bibinfo {author} {\bibfnamefont {Z.-H.}\ \bibnamefont {Zhu}},\ and\
  \bibinfo {author} {\bibfnamefont {C.}~\bibnamefont {Rosales-Guzm{\'{a}}n}},\
  }\bibfield  {title} {\bibinfo {title} {Classically entangled
  {I}nce{\textendash}{G}aussian modes},\ }\href
  {https://doi.org/10.1063/5.0011142} {\bibfield  {journal} {\bibinfo
  {journal} {Appl. Phys. Lett.}\ }\textbf {\bibinfo {volume} {116}},\ \bibinfo
  {pages} {221105} (\bibinfo {year} {2020})}\BibitemShut {NoStop}%
\bibitem [{\citenamefont {Sakpal}\ \emph {et~al.}(2018)\citenamefont {Sakpal},
  \citenamefont {Milione}, \citenamefont {Li}, \citenamefont {Nouri},
  \citenamefont {Shahoei}, \citenamefont {LaFave}, \citenamefont {Ashrafi},\
  and\ \citenamefont {MacFarlane}}]{sakpal2018stability}%
  \BibitemOpen
  \bibfield  {author} {\bibinfo {author} {\bibfnamefont {S.}~\bibnamefont
  {Sakpal}}, \bibinfo {author} {\bibfnamefont {G.}~\bibnamefont {Milione}},
  \bibinfo {author} {\bibfnamefont {M.-J.}\ \bibnamefont {Li}}, \bibinfo
  {author} {\bibfnamefont {M.}~\bibnamefont {Nouri}}, \bibinfo {author}
  {\bibfnamefont {H.}~\bibnamefont {Shahoei}}, \bibinfo {author} {\bibfnamefont
  {T.}~\bibnamefont {LaFave}}, \bibinfo {author} {\bibfnamefont
  {S.}~\bibnamefont {Ashrafi}},\ and\ \bibinfo {author} {\bibfnamefont
  {D.}~\bibnamefont {MacFarlane}},\ }\bibfield  {title} {\bibinfo {title}
  {Stability of ince{\textendash}gaussian beams in elliptical core few-mode
  fibers},\ }\href {https://doi.org/10.1364/ol.43.002656} {\bibfield  {journal}
  {\bibinfo  {journal} {Opt. Lett.}\ }\textbf {\bibinfo {volume} {43}},\
  \bibinfo {pages} {2656} (\bibinfo {year} {2018})}\BibitemShut {NoStop}%
\bibitem [{\citenamefont {Guti\'errez-Cuevas}\ \emph
  {et~al.}(2023)\citenamefont {Guti\'errez-Cuevas}, \citenamefont {O'Dell},
  \citenamefont {Dennis},\ and\ \citenamefont
  {Alonso}}]{gutierrezcuevas2023exactly}%
  \BibitemOpen
  \bibfield  {author} {\bibinfo {author} {\bibfnamefont {R.}~\bibnamefont
  {Guti\'errez-Cuevas}}, \bibinfo {author} {\bibfnamefont {D.~H.~J.}\
  \bibnamefont {O'Dell}}, \bibinfo {author} {\bibfnamefont {M.~R.}\
  \bibnamefont {Dennis}},\ and\ \bibinfo {author} {\bibfnamefont {M.~A.}\
  \bibnamefont {Alonso}},\ }\bibfield  {title} {\bibinfo {title} {Exactly
  solvable model behind {B}ose-{H}ubbard dimers, {I}nce-{G}auss beams, and
  aberrated optical cavities},\ }\href
  {https://doi.org/10.1103/PhysRevA.107.L031502} {\bibfield  {journal}
  {\bibinfo  {journal} {Phys. Rev. A}\ }\textbf {\bibinfo {volume} {107}},\
  \bibinfo {pages} {L031502} (\bibinfo {year} {2023})}\BibitemShut {NoStop}%
\bibitem [{\citenamefont {Dennis}(2006)}]{dennis2006rows}%
  \BibitemOpen
  \bibfield  {author} {\bibinfo {author} {\bibfnamefont {M.~R.}\ \bibnamefont
  {Dennis}},\ }\bibfield  {title} {\bibinfo {title} {Rows of optical vortices
  from elliptically perturbing a high-order beam},\ }\href
  {https://doi.org/10.1364/ol.31.001325} {\bibfield  {journal} {\bibinfo
  {journal} {Opt. Lett.}\ }\textbf {\bibinfo {volume} {31}},\ \bibinfo {pages}
  {1325} (\bibinfo {year} {2006})}\BibitemShut {NoStop}%
\bibitem [{\citenamefont {Woerdemann}\ \emph {et~al.}(2011)\citenamefont
  {Woerdemann}, \citenamefont {Alpmann},\ and\ \citenamefont
  {Denz}}]{woerdemann2011optical}%
  \BibitemOpen
  \bibfield  {author} {\bibinfo {author} {\bibfnamefont {M.}~\bibnamefont
  {Woerdemann}}, \bibinfo {author} {\bibfnamefont {C.}~\bibnamefont
  {Alpmann}},\ and\ \bibinfo {author} {\bibfnamefont {C.}~\bibnamefont
  {Denz}},\ }\bibfield  {title} {\bibinfo {title} {Optical assembly of
  microparticles into highly ordered structures using ince{\textendash}gaussian
  beams},\ }\href {https://doi.org/10.1063/1.3561770} {\bibfield  {journal}
  {\bibinfo  {journal} {Appl. Phys. Lett.}\ }\textbf {\bibinfo {volume} {98}},\
  \bibinfo {pages} {111101} (\bibinfo {year} {2011})}\BibitemShut {NoStop}%
\bibitem [{\citenamefont {Krenn}\ \emph {et~al.}(2013)\citenamefont {Krenn},
  \citenamefont {Fickler}, \citenamefont {Huber}, \citenamefont {Lapkiewicz},
  \citenamefont {Plick}, \citenamefont {Ramelow},\ and\ \citenamefont
  {Zeilinger}}]{krenn2013entangled}%
  \BibitemOpen
  \bibfield  {author} {\bibinfo {author} {\bibfnamefont {M.}~\bibnamefont
  {Krenn}}, \bibinfo {author} {\bibfnamefont {R.}~\bibnamefont {Fickler}},
  \bibinfo {author} {\bibfnamefont {M.}~\bibnamefont {Huber}}, \bibinfo
  {author} {\bibfnamefont {R.}~\bibnamefont {Lapkiewicz}}, \bibinfo {author}
  {\bibfnamefont {W.}~\bibnamefont {Plick}}, \bibinfo {author} {\bibfnamefont
  {S.}~\bibnamefont {Ramelow}},\ and\ \bibinfo {author} {\bibfnamefont
  {A.}~\bibnamefont {Zeilinger}},\ }\bibfield  {title} {\bibinfo {title}
  {Entangled singularity patterns of photons in ince-gauss modes},\ }\href
  {https://doi.org/10.1103/physreva.87.012326} {\bibfield  {journal} {\bibinfo
  {journal} {Phys. Rev. A}\ }\textbf {\bibinfo {volume} {87}},\ \bibinfo
  {pages} {012326} (\bibinfo {year} {2013})}\BibitemShut {NoStop}%
\bibitem [{\citenamefont {Plick}\ \emph {et~al.}(2013)\citenamefont {Plick},
  \citenamefont {Krenn}, \citenamefont {Fickler}, \citenamefont {Ramelow},\
  and\ \citenamefont {Zeilinger}}]{plick2013quantum}%
  \BibitemOpen
  \bibfield  {author} {\bibinfo {author} {\bibfnamefont {W.~N.}\ \bibnamefont
  {Plick}}, \bibinfo {author} {\bibfnamefont {M.}~\bibnamefont {Krenn}},
  \bibinfo {author} {\bibfnamefont {R.}~\bibnamefont {Fickler}}, \bibinfo
  {author} {\bibfnamefont {S.}~\bibnamefont {Ramelow}},\ and\ \bibinfo {author}
  {\bibfnamefont {A.}~\bibnamefont {Zeilinger}},\ }\bibfield  {title} {\bibinfo
  {title} {Quantum orbital angular momentum of elliptically symmetric light},\
  }\href {https://doi.org/10.1103/physreva.87.033806} {\bibfield  {journal}
  {\bibinfo  {journal} {Phys. Rev. A}\ }\textbf {\bibinfo {volume} {87}},\
  \bibinfo {pages} {033806} (\bibinfo {year} {2013})}\BibitemShut {NoStop}%
\bibitem [{\citenamefont {Dennis}\ and\ \citenamefont
  {Alonso}(2017)}]{dennis2017swings}%
  \BibitemOpen
  \bibfield  {author} {\bibinfo {author} {\bibfnamefont {M.~R.}\ \bibnamefont
  {Dennis}}\ and\ \bibinfo {author} {\bibfnamefont {M.~A.}\ \bibnamefont
  {Alonso}},\ }\bibfield  {title} {\bibinfo {title} {Swings and roundabouts:
  optical {P}oincar{\'{e}} spheres for polarization and {G}aussian beams},\
  }\href {https://doi.org/10.1098/rsta.2015.0441} {\bibfield  {journal}
  {\bibinfo  {journal} {Phil. Trans. R. Soc. A}\ }\textbf {\bibinfo {volume}
  {375}},\ \bibinfo {pages} {20150441} (\bibinfo {year} {2017})}\BibitemShut
  {NoStop}%
\bibitem [{\citenamefont {Malhotra}\ \emph {et~al.}(2018)\citenamefont
  {Malhotra}, \citenamefont {Guti{\'{e}}rrez-Cuevas}, \citenamefont {Hassett},
  \citenamefont {Dennis}, \citenamefont {Vamivakas},\ and\ \citenamefont
  {Alonso}}]{malhotra2018measuring}%
  \BibitemOpen
  \bibfield  {author} {\bibinfo {author} {\bibfnamefont {T.}~\bibnamefont
  {Malhotra}}, \bibinfo {author} {\bibfnamefont {R.}~\bibnamefont
  {Guti{\'{e}}rrez-Cuevas}}, \bibinfo {author} {\bibfnamefont {J.}~\bibnamefont
  {Hassett}}, \bibinfo {author} {\bibfnamefont {M.}~\bibnamefont {Dennis}},
  \bibinfo {author} {\bibfnamefont {A.}~\bibnamefont {Vamivakas}},\ and\
  \bibinfo {author} {\bibfnamefont {M.}~\bibnamefont {Alonso}},\ }\bibfield
  {title} {\bibinfo {title} {Measuring geometric phase without
  interferometry},\ }\href {https://doi.org/10.1103/physrevlett.120.233602}
  {\bibfield  {journal} {\bibinfo  {journal} {Phys. Rev. Lett.}\ }\textbf
  {\bibinfo {volume} {120}},\ \bibinfo {pages} {233602} (\bibinfo {year}
  {2018})}\BibitemShut {NoStop}%
\bibitem [{\citenamefont {Guti{\'{e}}rrez-Cuevas}\ \emph
  {et~al.}(2019)\citenamefont {Guti{\'{e}}rrez-Cuevas}, \citenamefont
  {Dennis},\ and\ \citenamefont {Alonso}}]{gutierrez-cuevas2019generalized}%
  \BibitemOpen
  \bibfield  {author} {\bibinfo {author} {\bibfnamefont {R.}~\bibnamefont
  {Guti{\'{e}}rrez-Cuevas}}, \bibinfo {author} {\bibfnamefont {M.~R.}\
  \bibnamefont {Dennis}},\ and\ \bibinfo {author} {\bibfnamefont {M.~A.}\
  \bibnamefont {Alonso}},\ }\bibfield  {title} {\bibinfo {title} {Generalized
  {G}aussian beams in terms of {J}ones vectors},\ }\href
  {https://doi.org/10.1088/2040-8986/ab2c52} {\bibfield  {journal} {\bibinfo
  {journal} {J. Opt.}\ }\textbf {\bibinfo {volume} {21}},\ \bibinfo {pages}
  {084001} (\bibinfo {year} {2019})}\BibitemShut {NoStop}%
\bibitem [{\citenamefont {Guti{\'{e}}rrez-Cuevas}\ \emph
  {et~al.}(2020)\citenamefont {Guti{\'{e}}rrez-Cuevas}, \citenamefont {Wadood},
  \citenamefont {Vamivakas},\ and\ \citenamefont
  {Alonso}}]{gutierrez-cuevas2020modal}%
  \BibitemOpen
  \bibfield  {author} {\bibinfo {author} {\bibfnamefont {R.}~\bibnamefont
  {Guti{\'{e}}rrez-Cuevas}}, \bibinfo {author} {\bibfnamefont {S.}~\bibnamefont
  {Wadood}}, \bibinfo {author} {\bibfnamefont {A.}~\bibnamefont {Vamivakas}},\
  and\ \bibinfo {author} {\bibfnamefont {M.}~\bibnamefont {Alonso}},\
  }\bibfield  {title} {\bibinfo {title} {Modal majorana sphere and hidden
  symmetries of structured-gaussian beams},\ }\href
  {https://doi.org/10.1103/physrevlett.125.123903} {\bibfield  {journal}
  {\bibinfo  {journal} {Phys. Rev. Lett.}\ }\textbf {\bibinfo {volume} {125}},\
  \bibinfo {pages} {123903} (\bibinfo {year} {2020})}\BibitemShut {NoStop}%
\bibitem [{\citenamefont {Guti{\'{e}}rrez-Cuevas}\ and\ \citenamefont
  {Alonso}(2020)}]{gutierrezcuevas2020platonic}%
  \BibitemOpen
  \bibfield  {author} {\bibinfo {author} {\bibfnamefont {R.}~\bibnamefont
  {Guti{\'{e}}rrez-Cuevas}}\ and\ \bibinfo {author} {\bibfnamefont {M.~A.}\
  \bibnamefont {Alonso}},\ }\bibfield  {title} {\bibinfo {title} {Platonic
  gaussian beams: wave and ray treatment},\ }\href
  {https://doi.org/10.1364/ol.405988} {\bibfield  {journal} {\bibinfo
  {journal} {Opt. Lett.}\ }\textbf {\bibinfo {volume} {45}},\ \bibinfo {pages}
  {6759} (\bibinfo {year} {2020})}\BibitemShut {NoStop}%
\bibitem [{\citenamefont {Schwarz}\ \emph {et~al.}(2004)\citenamefont
  {Schwarz}, \citenamefont {Bandres},\ and\ \citenamefont
  {Guti{\'{e}}rrez-Vega}}]{schwarz2004observation}%
  \BibitemOpen
  \bibfield  {author} {\bibinfo {author} {\bibfnamefont {U.~T.}\ \bibnamefont
  {Schwarz}}, \bibinfo {author} {\bibfnamefont {M.~A.}\ \bibnamefont
  {Bandres}},\ and\ \bibinfo {author} {\bibfnamefont {J.~C.}\ \bibnamefont
  {Guti{\'{e}}rrez-Vega}},\ }\bibfield  {title} {\bibinfo {title} {Observation
  of {I}nce{\textendash}{G}aussian modes in stable resonators},\ }\href
  {https://doi.org/10.1364/ol.29.001870} {\bibfield  {journal} {\bibinfo
  {journal} {Opt. Lett.}\ }\textbf {\bibinfo {volume} {29}},\ \bibinfo {pages}
  {1870} (\bibinfo {year} {2004})}\BibitemShut {NoStop}%
\bibitem [{\citenamefont {Sakurai}\ and\ \citenamefont
  {Napolitano}(2010)}]{sakurai2010modern}%
  \BibitemOpen
  \bibfield  {author} {\bibinfo {author} {\bibfnamefont {J.~J.}\ \bibnamefont
  {Sakurai}}\ and\ \bibinfo {author} {\bibfnamefont {J.~J.}\ \bibnamefont
  {Napolitano}},\ }\href
  {https://www.amazon.com/Modern-Quantum-Mechanics-2nd-Sakurai/dp/0805382917?SubscriptionId=AKIAIOBINVZYXZQZ2U3A&tag=chimbori05-20&linkCode=xm2&camp=2025&creative=165953&creativeASIN=0805382917}
  {\emph {\bibinfo {title} {Modern Quantum Mechanics (2nd Edition)}}}\
  (\bibinfo  {publisher} {Pearson},\ \bibinfo {year} {2010})\BibitemShut
  {NoStop}%
\bibitem [{\citenamefont {Padgett}\ and\ \citenamefont
  {Courtial}(1999)}]{padgett1999poincare}%
  \BibitemOpen
  \bibfield  {author} {\bibinfo {author} {\bibfnamefont {M.~J.}\ \bibnamefont
  {Padgett}}\ and\ \bibinfo {author} {\bibfnamefont {J.}~\bibnamefont
  {Courtial}},\ }\bibfield  {title} {\bibinfo {title} {Poincar\'{e}-sphere
  equivalent for light beams containing orbital angular momentum},\ }\href
  {https://doi.org/10.1364/ol.24.000430} {\bibfield  {journal} {\bibinfo
  {journal} {Opt. Lett.}\ }\textbf {\bibinfo {volume} {24}},\ \bibinfo {pages}
  {430} (\bibinfo {year} {1999})}\BibitemShut {NoStop}%
\bibitem [{\citenamefont {Calvo}(2005)}]{calvo2005wigner}%
  \BibitemOpen
  \bibfield  {author} {\bibinfo {author} {\bibfnamefont {G.~F.}\ \bibnamefont
  {Calvo}},\ }\bibfield  {title} {\bibinfo {title} {Wigner representation and
  geometric transformations of optical orbital angular momentum spatial
  modes},\ }\href {https://doi.org/10.1364/ol.30.001207} {\bibfield  {journal}
  {\bibinfo  {journal} {Opt. Lett.}\ }\textbf {\bibinfo {volume} {30}},\
  \bibinfo {pages} {1207} (\bibinfo {year} {2005})}\BibitemShut {NoStop}%
\bibitem [{\citenamefont {Gati}\ and\ \citenamefont
  {Oberthaler}(2007)}]{gati2007bosonic}%
  \BibitemOpen
  \bibfield  {author} {\bibinfo {author} {\bibfnamefont {R.}~\bibnamefont
  {Gati}}\ and\ \bibinfo {author} {\bibfnamefont {M.~K.}\ \bibnamefont
  {Oberthaler}},\ }\bibfield  {title} {\bibinfo {title} {A bosonic josephson
  junction},\ }\href {https://doi.org/10.1088/0953-4075/40/10/r01} {\bibfield
  {journal} {\bibinfo  {journal} {J. Phys. B: At., Mol. Opt. Phys.}\ }\textbf
  {\bibinfo {volume} {40}},\ \bibinfo {pages} {R61} (\bibinfo {year}
  {2007})}\BibitemShut {NoStop}%
\bibitem [{\citenamefont {Berry}\ and\ \citenamefont
  {Upstill}(1980)}]{berry1980iv}%
  \BibitemOpen
  \bibfield  {author} {\bibinfo {author} {\bibfnamefont {M.}~\bibnamefont
  {Berry}}\ and\ \bibinfo {author} {\bibfnamefont {C.}~\bibnamefont
  {Upstill}},\ }\bibfield  {title} {\bibinfo {title} {{IV} catastrophe optics:
  Morphologies of caustics and their diffraction patterns},\ }in\ \href
  {https://doi.org/10.1016/s0079-6638(08)70215-4} {\emph {\bibinfo {booktitle}
  {Prog. Opt.}}}\ (\bibinfo  {publisher} {Elsevier},\ \bibinfo {year} {1980})\
  pp.\ \bibinfo {pages} {257--346}\BibitemShut {NoStop}%
\bibitem [{\citenamefont {Nye}(1999)}]{nye1999natural}%
  \BibitemOpen
  \bibfield  {author} {\bibinfo {author} {\bibfnamefont {J.}~\bibnamefont
  {Nye}},\ }\href@noop {} {\emph {\bibinfo {title} {Natural Focusing and Fine
  Structure of Light: Caustics and Wave Dislocations}}}\ (\bibinfo  {publisher}
  {Institute of Physics Publishing: Bristol and Philadelphia},\ \bibinfo {year}
  {1999})\BibitemShut {NoStop}%
\bibitem [{\citenamefont {Kravtsov}\ and\ \citenamefont
  {Orlov}(1999)}]{kravtsov1999caustics}%
  \BibitemOpen
  \bibfield  {author} {\bibinfo {author} {\bibfnamefont {Y.~A.}\ \bibnamefont
  {Kravtsov}}\ and\ \bibinfo {author} {\bibfnamefont {Y.~I.}\ \bibnamefont
  {Orlov}},\ }\href@noop {} {\emph {\bibinfo {title} {Caustics, Catastrophes
  and Wave Fields}}}\ (\bibinfo  {publisher} {Springer},\ \bibinfo {year}
  {1999})\BibitemShut {NoStop}%
\bibitem [{\citenamefont {Alonso}\ and\ \citenamefont
  {Forbes}(2002)}]{alonso2002stable}%
  \BibitemOpen
  \bibfield  {author} {\bibinfo {author} {\bibfnamefont {M.}~\bibnamefont
  {Alonso}}\ and\ \bibinfo {author} {\bibfnamefont {G.}~\bibnamefont
  {Forbes}},\ }\bibfield  {title} {\bibinfo {title} {Stable aggregates of
  flexible elements give a stronger link between rays and waves},\ }\href
  {https://doi.org/10.1364/oe.10.000728} {\bibfield  {journal} {\bibinfo
  {journal} {Opt. Express}\ }\textbf {\bibinfo {volume} {10}},\ \bibinfo
  {pages} {728} (\bibinfo {year} {2002})}\BibitemShut {NoStop}%
\bibitem [{\citenamefont {Alonso}(2009)}]{alonso2009rays}%
  \BibitemOpen
  \bibfield  {author} {\bibinfo {author} {\bibfnamefont {M.~A.}\ \bibnamefont
  {Alonso}},\ }\bibfield  {title} {\bibinfo {title} {Rays and waves},\ }in\
  \href@noop {} {\emph {\bibinfo {booktitle} {Phase Space Optics: Fundamentals
  and Applications}}}\ (\bibinfo  {publisher} {McGraw Hill Professional},\
  \bibinfo {year} {2009})\ pp.\ \bibinfo {pages} {237--277}\BibitemShut
  {NoStop}%
\bibitem [{\citenamefont {Shen}\ \emph {et~al.}(2020)\citenamefont {Shen},
  \citenamefont {Yang}, \citenamefont {Naidoo}, \citenamefont {Fu},\ and\
  \citenamefont {Forbes}}]{shen2020structured}%
  \BibitemOpen
  \bibfield  {author} {\bibinfo {author} {\bibfnamefont {Y.}~\bibnamefont
  {Shen}}, \bibinfo {author} {\bibfnamefont {X.}~\bibnamefont {Yang}}, \bibinfo
  {author} {\bibfnamefont {D.}~\bibnamefont {Naidoo}}, \bibinfo {author}
  {\bibfnamefont {X.}~\bibnamefont {Fu}},\ and\ \bibinfo {author}
  {\bibfnamefont {A.}~\bibnamefont {Forbes}},\ }\bibfield  {title} {\bibinfo
  {title} {Structured ray-wave vector vortex beams in multiple degrees of
  freedom from a laser},\ }\href {https://doi.org/10.1364/optica.382994}
  {\bibfield  {journal} {\bibinfo  {journal} {Optica}\ }\textbf {\bibinfo
  {volume} {7}},\ \bibinfo {pages} {820} (\bibinfo {year} {2020})}\BibitemShut
  {NoStop}%
\bibitem [{\citenamefont {Abramochkin}\ and\ \citenamefont
  {Volostnikov}(2010)}]{abramochkin2010generalized}%
  \BibitemOpen
  \bibfield  {author} {\bibinfo {author} {\bibfnamefont {E.~G.}\ \bibnamefont
  {Abramochkin}}\ and\ \bibinfo {author} {\bibfnamefont {V.~G.}\ \bibnamefont
  {Volostnikov}},\ }\bibfield  {title} {\bibinfo {title} {Generalized
  hermite-laguerre-gauss beams},\ }\href
  {https://doi.org/10.3103/s1541308x10010036} {\bibfield  {journal} {\bibinfo
  {journal} {Phys. Wave Phenom.}\ }\textbf {\bibinfo {volume} {18}},\ \bibinfo
  {pages} {14} (\bibinfo {year} {2010})}\BibitemShut {NoStop}%
\bibitem [{\citenamefont {lin Chao}(1974)}]{chao1974high}%
  \BibitemOpen
  \bibfield  {author} {\bibinfo {author} {\bibfnamefont {S.}~\bibnamefont {lin
  Chao}},\ }\emph {\bibinfo {title} {High order transverse modes in an
  astigmatic cavity}},\ \href@noop {} {Ph.D. thesis},\ \bibinfo  {school}
  {University of Rochester} (\bibinfo {year} {1974})\BibitemShut {NoStop}%
\bibitem [{\citenamefont {Abbena}\ \emph {et~al.}(2017)\citenamefont {Abbena},
  \citenamefont {Salamon},\ and\ \citenamefont {Gray}}]{abbena2017modern}%
  \BibitemOpen
  \bibfield  {author} {\bibinfo {author} {\bibfnamefont {E.}~\bibnamefont
  {Abbena}}, \bibinfo {author} {\bibfnamefont {S.}~\bibnamefont {Salamon}},\
  and\ \bibinfo {author} {\bibfnamefont {A.}~\bibnamefont {Gray}},\ }\href@noop
  {} {\emph {\bibinfo {title} {Modern differential geometry of curves and
  surfaces with Mathematica}}}\ (\bibinfo  {publisher} {CRC press},\ \bibinfo
  {year} {2017})\BibitemShut {NoStop}%
\bibitem [{\citenamefont {Graefe}\ \emph {et~al.}(2014)\citenamefont {Graefe},
  \citenamefont {Korsch},\ and\ \citenamefont
  {Strzys}}]{graefe2014bosehubbard}%
  \BibitemOpen
  \bibfield  {author} {\bibinfo {author} {\bibfnamefont {E.-M.}\ \bibnamefont
  {Graefe}}, \bibinfo {author} {\bibfnamefont {H.~J.}\ \bibnamefont {Korsch}},\
  and\ \bibinfo {author} {\bibfnamefont {M.~P.}\ \bibnamefont {Strzys}},\
  }\bibfield  {title} {\bibinfo {title} {Bose{\textendash}{H}ubbard dimers,
  {V}iviani's windows and pendulum dynamics},\ }\href
  {https://doi.org/10.1088/1751-8113/47/8/085304} {\bibfield  {journal}
  {\bibinfo  {journal} {J. Phys. A: Math. Theor.}\ }\textbf {\bibinfo {volume}
  {47}},\ \bibinfo {pages} {085304} (\bibinfo {year} {2014})}\BibitemShut
  {NoStop}%
\bibitem [{\citenamefont {Morales}\ \emph {et~al.}(2017)\citenamefont
  {Morales}, \citenamefont {Rodr{\'{\i}}guez-Lara},\ and\ \citenamefont
  {Malomed}}]{morales2017polarization}%
  \BibitemOpen
  \bibfield  {author} {\bibinfo {author} {\bibfnamefont {J.~D.~H.}\
  \bibnamefont {Morales}}, \bibinfo {author} {\bibfnamefont {B.~M.}\
  \bibnamefont {Rodr{\'{\i}}guez-Lara}},\ and\ \bibinfo {author} {\bibfnamefont
  {B.~A.}\ \bibnamefont {Malomed}},\ }\bibfield  {title} {\bibinfo {title}
  {Polarization dynamics in twisted fiber amplifiers: a non-hermitian nonlinear
  dimer model},\ }\href {https://doi.org/10.1364/ol.42.004402} {\bibfield
  {journal} {\bibinfo  {journal} {Opt. Lett.}\ }\textbf {\bibinfo {volume}
  {42}},\ \bibinfo {pages} {4402} (\bibinfo {year} {2017})}\BibitemShut
  {NoStop}%
\bibitem [{\citenamefont {Azzam}(2000)}]{azzam2000poincare}%
  \BibitemOpen
  \bibfield  {author} {\bibinfo {author} {\bibfnamefont {R.~M.~A.}\
  \bibnamefont {Azzam}},\ }\bibfield  {title} {\bibinfo {title} {Poincar\'{e}
  sphere representation of the fixed-polarizer rotating-retarder optical
  system},\ }\href {https://doi.org/10.1364/JOSAA.17.002105} {\bibfield
  {journal} {\bibinfo  {journal} {J. Opt. Soc. Am. A}\ }\textbf {\bibinfo
  {volume} {17}},\ \bibinfo {pages} {2105} (\bibinfo {year}
  {2000})}\BibitemShut {NoStop}%
\bibitem [{\citenamefont {Blum}(1967)}]{blum1967transformation}%
  \BibitemOpen
  \bibfield  {author} {\bibinfo {author} {\bibfnamefont {H.}~\bibnamefont
  {Blum}},\ }\bibfield  {title} {\bibinfo {title} {A transformation for
  extracting new descriptors of shape},\ }in\ \href@noop {} {\emph {\bibinfo
  {booktitle} {Models for the Perception of Speech and Visual Form}}},\
  \bibinfo {editor} {edited by\ \bibinfo {editor} {\bibfnamefont
  {W.}~\bibnamefont {Wathen-Dunn}}}\ (\bibinfo  {publisher} {MIT Press},\
  \bibinfo {year} {1967})\ p.\ \bibinfo {pages} {362–380}\BibitemShut
  {NoStop}%
\bibitem [{\citenamefont {Ogilvy}(1990)}]{ogilvy1990excursions}%
  \BibitemOpen
  \bibfield  {author} {\bibinfo {author} {\bibfnamefont {C.~S.}\ \bibnamefont
  {Ogilvy}},\ }\href@noop {} {\emph {\bibinfo {title} {Excursions in
  geometry}}}\ (\bibinfo  {publisher} {Courier Corporation},\ \bibinfo {year}
  {1990})\BibitemShut {NoStop}%
\bibitem [{\citenamefont {Mittag}\ and\ \citenamefont
  {Stephen}(1992)}]{mittag1992conformal}%
  \BibitemOpen
  \bibfield  {author} {\bibinfo {author} {\bibfnamefont {L.}~\bibnamefont
  {Mittag}}\ and\ \bibinfo {author} {\bibfnamefont {M.~J.}\ \bibnamefont
  {Stephen}},\ }\bibfield  {title} {\bibinfo {title} {Conformal transformations
  and the application of complex variables in mechanics and quantum
  mechanics},\ }\href {https://doi.org/10.1119/1.16948} {\bibfield  {journal}
  {\bibinfo  {journal} {Am. J. Phys}\ }\textbf {\bibinfo {volume} {60}},\
  \bibinfo {pages} {207} (\bibinfo {year} {1992})}\BibitemShut {NoStop}%
\bibitem [{\citenamefont {Alem{\'{a}}n-Casta{\~{n}}eda}\ and\ \citenamefont
  {Alonso}(2019)}]{aleman-castaneda2019study}%
  \BibitemOpen
  \bibfield  {author} {\bibinfo {author} {\bibfnamefont {L.~A.}\ \bibnamefont
  {Alem{\'{a}}n-Casta{\~{n}}eda}}\ and\ \bibinfo {author} {\bibfnamefont
  {M.~A.}\ \bibnamefont {Alonso}},\ }\bibfield  {title} {\bibinfo {title}
  {Study of reflectors for illumination via conformal maps},\ }\href
  {https://doi.org/10.1364/ol.44.003809} {\bibfield  {journal} {\bibinfo
  {journal} {Opt. Lett.}\ }\textbf {\bibinfo {volume} {44}},\ \bibinfo {pages}
  {3809} (\bibinfo {year} {2019})}\BibitemShut {NoStop}%
\bibitem [{\citenamefont {Zor}\ and\ \citenamefont
  {Kay}(1996)}]{zor1996globally}%
  \BibitemOpen
  \bibfield  {author} {\bibinfo {author} {\bibfnamefont {D.}~\bibnamefont
  {Zor}}\ and\ \bibinfo {author} {\bibfnamefont {K.~G.}\ \bibnamefont {Kay}},\
  }\bibfield  {title} {\bibinfo {title} {Globally uniform semiclassical
  expressions for time-independent wave functions},\ }\href
  {https://doi.org/10.1103/physrevlett.76.1990} {\bibfield  {journal} {\bibinfo
   {journal} {Phys. Rev. Lett.}\ }\textbf {\bibinfo {volume} {76}},\ \bibinfo
  {pages} {1990} (\bibinfo {year} {1996})}\BibitemShut {NoStop}%
\bibitem [{\citenamefont {Alonso}\ and\ \citenamefont
  {Forbes}(2001)}]{alonso2001using}%
  \BibitemOpen
  \bibfield  {author} {\bibinfo {author} {\bibfnamefont {M.~A.}\ \bibnamefont
  {Alonso}}\ and\ \bibinfo {author} {\bibfnamefont {G.~W.}\ \bibnamefont
  {Forbes}},\ }\bibfield  {title} {\bibinfo {title} {Using rays better {II} ray
  families to match prescribed wave fields},\ }\href
  {https://doi.org/10.1364/josaa.18.001146} {\bibfield  {journal} {\bibinfo
  {journal} {J. Opt. Soc. Am. A}\ }\textbf {\bibinfo {volume} {18}},\ \bibinfo
  {pages} {1146} (\bibinfo {year} {2001})}\BibitemShut {NoStop}%
\bibitem [{\citenamefont {Kay}(1999)}]{kay1999exact}%
  \BibitemOpen
  \bibfield  {author} {\bibinfo {author} {\bibfnamefont {K.~G.}\ \bibnamefont
  {Kay}},\ }\bibfield  {title} {\bibinfo {title} {Exact wave functions for the
  coulomb problem from classical orbits},\ }\href
  {https://doi.org/10.1103/physrevlett.83.5190} {\bibfield  {journal} {\bibinfo
   {journal} {Phys. Rev. Lett.}\ }\textbf {\bibinfo {volume} {83}},\ \bibinfo
  {pages} {5190} (\bibinfo {year} {1999})}\BibitemShut {NoStop}%
\bibitem [{\citenamefont {Kay}(2001)}]{kay2001exact}%
  \BibitemOpen
  \bibfield  {author} {\bibinfo {author} {\bibfnamefont {K.~G.}\ \bibnamefont
  {Kay}},\ }\bibfield  {title} {\bibinfo {title} {Exact wave functions from
  classical orbits: The isotropic harmonic oscillator and semiclassical
  applications},\ }\href {https://doi.org/10.1103/physreva.63.042110}
  {\bibfield  {journal} {\bibinfo  {journal} {Phys. Rev. A}\ }\textbf {\bibinfo
  {volume} {63}},\ \bibinfo {pages} {042110} (\bibinfo {year}
  {2001})}\BibitemShut {NoStop}%
\bibitem [{\citenamefont {Radcliffe}(1971)}]{radcliffe1971some}%
  \BibitemOpen
  \bibfield  {author} {\bibinfo {author} {\bibfnamefont {J.~M.}\ \bibnamefont
  {Radcliffe}},\ }\bibfield  {title} {\bibinfo {title} {Some properties of
  coherent spin states},\ }\href {https://doi.org/10.1088/0305-4470/4/3/009}
  {\bibfield  {journal} {\bibinfo  {journal} {J. Phys. A: Gen. Phys.}\ }\textbf
  {\bibinfo {volume} {4}},\ \bibinfo {pages} {313} (\bibinfo {year}
  {1971})}\BibitemShut {NoStop}%
\bibitem [{\citenamefont {Arecchi}\ \emph {et~al.}(1972)\citenamefont
  {Arecchi}, \citenamefont {Courtens}, \citenamefont {Gilmore},\ and\
  \citenamefont {Thomas}}]{arecchi1972atomic}%
  \BibitemOpen
  \bibfield  {author} {\bibinfo {author} {\bibfnamefont {F.~T.}\ \bibnamefont
  {Arecchi}}, \bibinfo {author} {\bibfnamefont {E.}~\bibnamefont {Courtens}},
  \bibinfo {author} {\bibfnamefont {R.}~\bibnamefont {Gilmore}},\ and\ \bibinfo
  {author} {\bibfnamefont {H.}~\bibnamefont {Thomas}},\ }\bibfield  {title}
  {\bibinfo {title} {Atomic coherent states in quantum optics},\ }\href
  {https://doi.org/10.1103/physreva.6.2211} {\bibfield  {journal} {\bibinfo
  {journal} {Phys. Rev. A}\ }\textbf {\bibinfo {volume} {6}},\ \bibinfo {pages}
  {2211} (\bibinfo {year} {1972})}\BibitemShut {NoStop}%
\bibitem [{\citenamefont {Perelomov}(1972)}]{perelomov1972coherent}%
  \BibitemOpen
  \bibfield  {author} {\bibinfo {author} {\bibfnamefont {A.~M.}\ \bibnamefont
  {Perelomov}},\ }\bibfield  {title} {\bibinfo {title} {Coherent states for
  arbitrary lie group},\ }\href {https://doi.org/10.1007/bf01645091} {\bibfield
   {journal} {\bibinfo  {journal} {Commun. Math. Phys.}\ }\textbf {\bibinfo
  {volume} {26}},\ \bibinfo {pages} {222} (\bibinfo {year} {1972})}\BibitemShut
  {NoStop}%
\bibitem [{\citenamefont {Forbes}\ and\ \citenamefont
  {Alonso}(2023)}]{forbes2023asymptotic}%
  \BibitemOpen
  \bibfield  {author} {\bibinfo {author} {\bibfnamefont {G.~W.}\ \bibnamefont
  {Forbes}}\ and\ \bibinfo {author} {\bibfnamefont {M.~A.}\ \bibnamefont
  {Alonso}},\ }\bibfield  {title} {\bibinfo {title} {Asymptotic expansions for
  field moments of bound states},\ }\href
  {https://doi.org/10.1088/1751-8121/acd310} {\bibfield  {journal} {\bibinfo
  {journal} {J. Phys. A: Math. Theor.}\ }\textbf {\bibinfo {volume} {56}},\
  \bibinfo {pages} {244001} (\bibinfo {year} {2023})}\BibitemShut {NoStop}%
\bibitem [{\citenamefont {Gutiérrez-Cuevas}()}]{repo}%
  \BibitemOpen
  \bibfield  {author} {\bibinfo {author} {\bibfnamefont {R.}~\bibnamefont
  {Gutiérrez-Cuevas}},\ }\href@noop {} {\bibinfo {title} {Incegauss}},\
  \bibinfo {howpublished}
  {\url{https://github.com/rodguti90/InceGauss}}\BibitemShut {NoStop}%
\end{thebibliography}%

\end{document}